\documentclass[10pt,a4paper]{article}
\usepackage[margin=21mm,top=21mm,bottom=21mm,headheight=13pt,headsep=7mm]{geometry}
\usepackage{lmodern,amsmath,graphicx,booktabs,caption,xcolor,hyperref,placeins}
\usepackage{iftex}
\ifXeTeX
\DeclareFontFamily{TU}{artss}{}
\DeclareFontShape{TU}{artss}{m}{n}{<-> \UnicodeFontFile{lmsans10-regular}{mapping=tex-text;}}{}
\DeclareFontShape{TU}{artss}{m}{it}{<-> \UnicodeFontFile{lmsans10-regular}{mapping=tex-text;slant=.16;}}{}
\IfFileExists{lmsans10-bold.otf}
 {\DeclareFontShape{TU}{artss}{b}{n}{<-> \UnicodeFontFile{lmsans10-bold}{mapping=tex-text;}}{}}%
 {\DeclareFontShape{TU}{artss}{b}{n}{<-> \UnicodeFontFile{lmsans10-regular}{mapping=tex-text;embolden=2;}}{}}%
\DeclareFontShape{TU}{artss}{bx}{n}{<->ssub*artss/b/n}{}

\else
  \ifPDFTeX
    \usepackage[T1]{fontenc}
    \DeclareUnicodeCharacter{0393}{\ensuremath{\Gamma}}
    \DeclareUnicodeCharacter{00B1}{\ensuremath{\pm}}
  \fi
\fi
\definecolor{ink}{HTML}{1B2930}
\definecolor{accent}{HTML}{176367}
\hypersetup{colorlinks=true,urlcolor=accent,linkcolor=accent,bookmarksnumbered=false}
\makeatletter
\renewcommand\section{\@startsection{section}{1}{0pt}{10pt plus 2pt minus 2pt}{4pt}{\sffamily\bfseries\large\color{ink}}}
\renewcommand\subsection{\@startsection{subsection}{2}{0pt}{7pt plus 2pt minus 1pt}{2pt}{\sffamily\bfseries\normalsize\color{ink}}}
\makeatother
\makeatletter
\def\ps@articlehead{\def\@oddhead{\normalfont\footnotesize\sffamily\rightmark\hfil\thepage}\let\@evenhead\@oddhead\let\@oddfoot\@empty\let\@evenfoot\@empty}
\makeatother
\begin{document}
\color{ink}

\markright{}
\thispagestyle{plain}
\hypersetup{pdftitle={El Agente en acci\'on: Closed-loop prediction of longer molecular qubit lifetimes}}
{\raggedright\sffamily\bfseries\fontsize{20}{23}\selectfont El Agente en acci\'on: Closed-loop prediction of longer molecular qubit lifetimes\par}\vspace{5pt}
\hypersetup{pdfauthor={Ignacio Gustin, Kahn Shah, Austin Cheng, Juan B. Pérez-Sánchez, Adem Chlyah, Alán Aspuru-Guzik, Varinia Bernales}}
{\raggedright\sffamily\bfseries\normalsize \mbox{Ignacio Gustin\textsuperscript{1,2}}, \mbox{Kahn Shah\textsuperscript{3}}, \mbox{Austin Cheng\textsuperscript{1,2,3,4}}, \mbox{Juan B. Pérez-Sánchez\textsuperscript{2}}, \mbox{Adem Chlyah\textsuperscript{5}},  \mbox{Alán Aspuru-Guzik\textsuperscript{1,2,3,4,6,7,8,9,10,*}},\mbox{Varinia Bernales\textsuperscript{2,3,9,*}}\par}
{\raggedright\footnotesize\setlength{\parskip}{1pt}
\hangindent=1.3em\hangafter=1\noindent\makebox[1.3em][l]{\textsuperscript{1}}NVIDIA, 431 King St. W \#6th, Toronto, ON M5V 1K4, Canada\par
\hangindent=1.3em\hangafter=1\noindent\makebox[1.3em][l]{\textsuperscript{2}}Department of Chemistry, University of Toronto, 80 St. George St., Toronto, ON M5S 3H6, Canada\par
\hangindent=1.3em\hangafter=1\noindent\makebox[1.3em][l]{\textsuperscript{3}}Department of Computer Science, University of Toronto, 40 St. George St., Toronto, ON M5S 2E4, Canada\par
\hangindent=1.3em\hangafter=1\noindent\makebox[1.3em][l]{\textsuperscript{4}}Vector Institute for Artificial Intelligence, W1140-108 College St., Schwartz Reisman Innovation Campus, Toronto, ON M5G 0C6, Canada\par
\hangindent=1.3em\hangafter=1\noindent\makebox[1.3em][l]{\textsuperscript{5}}Department of Mathematical and Computational Sciences, University of Toronto Mississauga, Mississauga, ON, Canada\newline
3359 Mississauga Rd, Mississauga, ON L5L 1C6\par
\hangindent=1.3em\hangafter=1\noindent\makebox[1.3em][l]{\textsuperscript{6}}Department of Materials Science \& Engineering, University of Toronto, 184 College St., Toronto, ON M5S 3E4, Canada\par
\hangindent=1.3em\hangafter=1\noindent\makebox[1.3em][l]{\textsuperscript{7}}Department of Chemical Engineering \& Applied Chemistry, University of Toronto, 200 College St., Toronto, ON M5S 3E5, Canada\par
\hangindent=1.3em\hangafter=1\noindent\makebox[1.3em][l]{\textsuperscript{8}}Acceleration Consortium, 700 University Ave., Toronto, ON M7A 2S4, Canada\par
\hangindent=1.3em\hangafter=1\noindent\makebox[1.3em][l]{\textsuperscript{9}}Canadian Institute for Advanced Research (CIFAR), 661 University Ave., Toronto, ON M5G 1M1, Canada\par
\hangindent=1.3em\hangafter=1\noindent\makebox[1.3em][l]{\textsuperscript{10}}Institute of Medical Science, 1 King’s College Circle, Medical Sciences Building, Room 2374, Toronto, ON M5S 1A8, Canada\par
\hangindent=1.3em\hangafter=1\noindent\makebox[1.3em][l]{\textsuperscript{*}}Corresponding authors: alan@aspuru.com, varinia@bernales.org\par
}
\pdfbookmark[1]{Abstract}{secAbstract}
\section*{Abstract}
Energy transfer from an electron spin to vibrations of the surrounding crystal (phonons) limits how long a molecular qubit can retain quantum information. Over five weeks, a team of supervised agents tested ways to slow this loss at room temperature. A discovery agent proposed chemical changes and predicted how each would affect spin relaxation. Specialist agents calculated crystal and electronic structures and modeled spin relaxation times. The discovery agent used their results to choose the next test. The team tested a benchmark vanadyl qubit with a measured spin–lattice relaxation time of 3.37 \(\mu\)s at 293 K in a magnetically diluted crystal. A predicted gain from adding fluorine to a neighbouring ion prompted the team to try chlorine and bromine at the same site. The spin-bearing molecule folded more in the bromine salt than in the chlorine salt, and its spin relaxed faster when the model allowed vibrations farther apart in frequency to contribute. The discovery agent chose the iodine analogue to test whether this trend extended to a heavier halogen and predicted that, under the same model, iodine would relax faster than the parent. The calculation gave that faster relaxation and, when the model favoured vibrations close in frequency, a longer lifetime for iodine. At phonon-pair broadenings of 3, 5 and 7 cm\(^{-1}\), substituting iodine or fluorine on the counterion extended the benchmark qubit’s calculated room-temperature spin lifetime by 22--53\% with iodine and 21--34\% with fluorine. Molecular qubits are a first application of this infrastructure. By changing the specialist calculations, the same discovery agent could explore how substitutions tune the colour of light from molecular quantum emitters, which solvents promote ion transport in battery electrolytes, or how pore chemistry controls water uptake in porous frameworks.
\par
\smallskip{\color{accent}\rule{19mm}{.45pt}}\par\smallskip
The ability to vary the structure of a molecule one chemical group at a time gives chemists a vast space in which to design materials for quantum applications. Molecular spin qubits illustrate what this freedom has already made possible. In these molecules, an electron spin can encode quantum information, and chemists have identified structures in which microwaves can manipulate the spin at room temperature\textsuperscript{\hyperlink{ref1}{1}} or light can prepare and read its state\textsuperscript{\hyperlink{ref2}{2}}. Using a spin to store information also requires that it retain its state after preparation. Here chemical intuition offers less guidance. Given a molecule whose spin is already long-lived, \textit{how should its structure or surroundings be changed to preserve that state for longer?}
\par
An experiment on a vanadyl qubit shows how far from the spin a consequential change can be made. In this salt, the spin resides on a negatively charged vanadyl complex, while two neighbouring counterions balance its charge. Replacing hydrogen with deuterium on the counterions alone increased the measured spin–lattice relaxation time, $T_1$, from 3.37 to 4.03 \(\mu\)s at 293 K in magnetically diluted crystals\textsuperscript{\hyperlink{ref3}{3}}. $T_1$ is the time over which the spin exchanges energy with its surroundings and its populations return to equilibrium. The surroundings can also affect phase memory, as studies in different hosts\textsuperscript{\hyperlink{ref4}{4}} and with different concentrations of nuclear spins\textsuperscript{\hyperlink{ref5}{5},\,\hyperlink{ref6}{6}} illustrate. Measurements of phase memory\textsuperscript{\hyperlink{ref7}{7},\,\hyperlink{ref8}{8}} also reflect dephasing. Phase coherence can decay even while the populations remain fixed\textsuperscript{\hyperlink{ref9}{9}}. In this paper, we focus on energy relaxation because it sets an upper bound of $2T_1$ on the coherence time and because the vanadyl experiment gives us a chemical starting point for extending it.
\par
The deuterium experiment identifies the counterion as a place to modify the qubit. To predict what another substitution will do, however, we have to follow its effects beyond the site of the change. A group that occupies more space on the counterion can alter how neighbouring ions pack around the vanadyl complex. If a new contact bends a ligand toward the metal, it changes both how the ligand moves\textsuperscript{\hyperlink{ref10}{10},\,\hyperlink{ref11}{11},\,\hyperlink{ref12}{12}} and how its motion perturbs the electronic states that carry the spin\textsuperscript{\hyperlink{ref13}{13},\,\hyperlink{ref14}{14},\,\hyperlink{ref15}{15},\,\hyperlink{ref16}{16}}. Meanwhile, the same change in packing can shift the frequencies of vibrations involving the surrounding ions. These effects meet in the relaxation process itself. A pair of vibrations can carry energy away from the spin when the difference between their energies matches the spin transition, and its contribution depends on how strongly those motions couple to the electronic states. A substitution can therefore weaken a vibration’s coupling to the spin while bringing other pairs closer to resonance. Whether the spin lives longer depends on the combined effect of these changes across the crystal.
\par
Tracing these competing pathways through a single chemical substitution requires expertise in crystal packing, electronic structure and spin dynamics. Setting up the simulations poses a second challenge because each calculation depends on choices made in the preceding one. For a salt with an unknown structure, several plausible packings must be compared before periodic density functional theory can refine the most promising structures and calculate their vibrations. Those vibrations determine how to displace the vanadyl complex in multireference electronic-structure calculations with ORCA. The resulting couplings must then be matched to the correct vibrational modes and electronic states in the spin-relaxation equations. A longer lifetime is evidence for a proposed design rule only if the calculations also support the physical explanation for the gain. Such an explanation can guide the next substitution, much as an energy-transfer hypothesis redirected a search for photostable molecules toward the solvent\textsuperscript{\hyperlink{ref17}{17}}.
\par

\par
Agents built on large language models have begun to take part in such searches, planning reactions on laboratory robots\textsuperscript{\hyperlink{ref18}{18}}, screening metal–organic frameworks\textsuperscript{\hyperlink{ref19}{19}} and proposing drug candidates\textsuperscript{\hyperlink{ref20}{20},\,\hyperlink{ref21}{21}} and nanobodies\textsuperscript{\hyperlink{ref22}{22}} that were then tested in experiments. Our groups have built the El Agente family one domain at a time: quantum chemistry\textsuperscript{\hyperlink{ref23}{23},\,\hyperlink{ref24}{24}}, molecular editing\textsuperscript{\hyperlink{ref25}{25}}, solid-state simulation\textsuperscript{\hyperlink{ref26}{26}}, machine-learned interatomic potentials\textsuperscript{\hyperlink{ref27}{27}}, and quantum simulation and dynamics\textsuperscript{\hyperlink{ref28}{28},\,\hyperlink{ref29}{29}}, with uses from quantum computing\textsuperscript{\hyperlink{ref28}{28}} to drug discovery\textsuperscript{\hyperlink{ref30}{30}}. Each agent was first benchmarked against problems with known answers. Quntur, which computed the electronic couplings in this work, scored 90.5 out of 100 on 17 expert-graded quantum chemistry tasks, against 61.7 for Claude Code in the same ORCA environment, each averaged over five independent repetitions\textsuperscript{\hyperlink{ref24}{24}}; in active-space selection for 32 singlet excitations, it lowered the mean absolute error against CASPT2 references by 48\% relative to a fixed protocol\textsuperscript{\hyperlink{ref31}{31}}. El Agente Cuántico, which formulated the relaxation model, averaged 97 to 100 out of 100 on five quantum-simulation tasks, each run ten times from the same prompt and graded by experts\textsuperscript{\hyperlink{ref28}{28}}. Estructural, which built the starting molecules, edits three-dimensional structures by substituting atoms and functional groups, exchanging ligands and controlling stereochemistry\textsuperscript{\hyperlink{ref25}{25}}, and El Agente Sólido scored 97.9\% on rubrics written by computational chemists across seven exercises, each repeated at least ten times\textsuperscript{\hyperlink{ref26}{26}}.
\par
This combination of specialist judgment and interdependent calculations makes molecular spin qubits a natural test for a multi-agent approach. Computational agents have linked crystal generation to experimental screening of metal–organic frameworks\textsuperscript{\hyperlink{ref19}{19}}, executed electronic-structure workflows\textsuperscript{\hyperlink{ref23}{23}}, selected calculation protocols from the literature\textsuperscript{\hyperlink{ref24}{24}} and translated physical models into simulations\textsuperscript{\hyperlink{ref28}{28}}. El Agente Sólido combines periodic DFT and phonon calculations\textsuperscript{\hyperlink{ref26}{26}}, and El Agente Forjador builds specialist agents for individual quantum simulation tasks\textsuperscript{\hyperlink{ref29}{29}}. Here, specialist agents set up the calculations for each salt and checked the derivation of the relaxation rates. A discovery agent compared the calculated lifetime and the pathways responsible for it with its recorded forecast, then used that comparison to propose the next chemical test (Fig. 1). In our five-week search, a ligand change weakened the targeted relaxation pathway, yet the spin relaxed faster overall. That result prompted the team to change the surrounding ions. The discovery agent chose iodine to test its prediction that the substitution would shorten the spin lifetime at broad phonon linewidth. The calculation confirmed that forecast at broad phonon linewidth and revealed a longer lifetime at narrow linewidth.
\par
\pdfbookmark[1]{Results}{secResults}
\section*{Results}
\subsection*{The agent network for molecular qubit discovery}
The discovery agent coordinated four specialist multi-agent systems. Estructural built the starting molecules, Quntur calculated electronic couplings, Cuántico formulated the relaxation model and Teórica checked its equations and physical limits. A hypothesis-generation agent helped guide the search by proposing physical mechanisms that the team could test through chemical substitutions. A literature specialist used Undermind to search for relevant prior work, including reported crystal structures for proposed counterion substitutions. Drawing on the literature and earlier calculations, the hypothesis agent predicted how a proposed change would affect spin relaxation and specified what result would challenge its explanation. The discovery agent recorded the forecast before the calculation began (\hyperlink{agentarchitecture.2}{Fig. 1a}). Specialist agents then calculated the crystal structure, electronic response and spin lifetime of the proposed salt. A separate numerical validator recomputed the lifetime from the submitted vibrations and couplings. The discovery agent used that result to decide which chemical change to test next, with the investigator approving extensions to the search.
\par
El Agente Estructural\textsuperscript{\hyperlink{ref25}{25}} generated the initial molecular structures for the proposed salts. For salts without a deposited crystal structure, we then predicted how the ions would pack. We used CLARI, a generative model for molecular crystal structure prediction\textsuperscript{\hyperlink{ref32}{32}}. Given the atom types and bonds of the constituent ions, CLARI uses a learned flow to generate the atomic positions and lattice vectors of a crystal unit cell together. Repeated sampling produced alternative packings of the same salt, yielding 6,798 candidate packings across 22 counterions. To screen these packings at low computational cost, we used the Universal Model for Atoms (UMA), a machine-learned interatomic potential\textsuperscript{\hyperlink{ref33}{33}}. We used the forces predicted by UMA to relax each trial structure, then ranked the relaxed packings of each salt by estimated free energy at 300 K. Up to 15 low-energy packings per salt advanced to periodic density functional theory (DFT) refinement in CP2K, which also supplied the vibrations of the selected structure. Quntur\textsuperscript{\hyperlink{ref24}{24}} used multireference calculations in ORCA to determine how atomic displacements changed the energies and couplings of the electronic spin states. Cuántico\textsuperscript{\hyperlink{ref28}{28}} formulated the spin Hamiltonian and relaxation-rate equations in terms of the electronic couplings and vibrational modes obtained from the preceding calculations. La Agente Teórica used symbolic calculations in Python (SymPy) and numerical simulations to check the derivations, verify the consistent use of parameters and test whether the equations reproduced the expected behaviour in known physical limits. The validator also tested whether the vibration targeted in the parent could be identified in the modified salt.
\par
Figure 1b shows when the discovery agent was active over the five-week search. From 16 October to 19 November 2026, 16 salt designs reached DFT. Fourteen received lifetime calculations, with two ordered structures for each of two designs. This gave 16 candidate crystal models plus the parent. The campaign used 673,419 allocated CPU core-hours including the pre-launch parent baseline, with phonons accounting for the largest share (Fig. 1c). Crystal prediction and related tests used 141 GPU-hours (Supplementary Tables S8--S9). The agents processed 1.87 billion input tokens, including cached prompt prefixes, and generated 12.0 million output tokens. The Methods detail the packing selection and relaxation procedures, and the Supplementary Information gives the agent activity and computational cost by stage.
\par
\begin{figure}[!htbp]
\pdfbookmark[2]{Fig. 1. The five-week agent campaign}{agentarchitecture}
\centering\includegraphics[width=\linewidth]{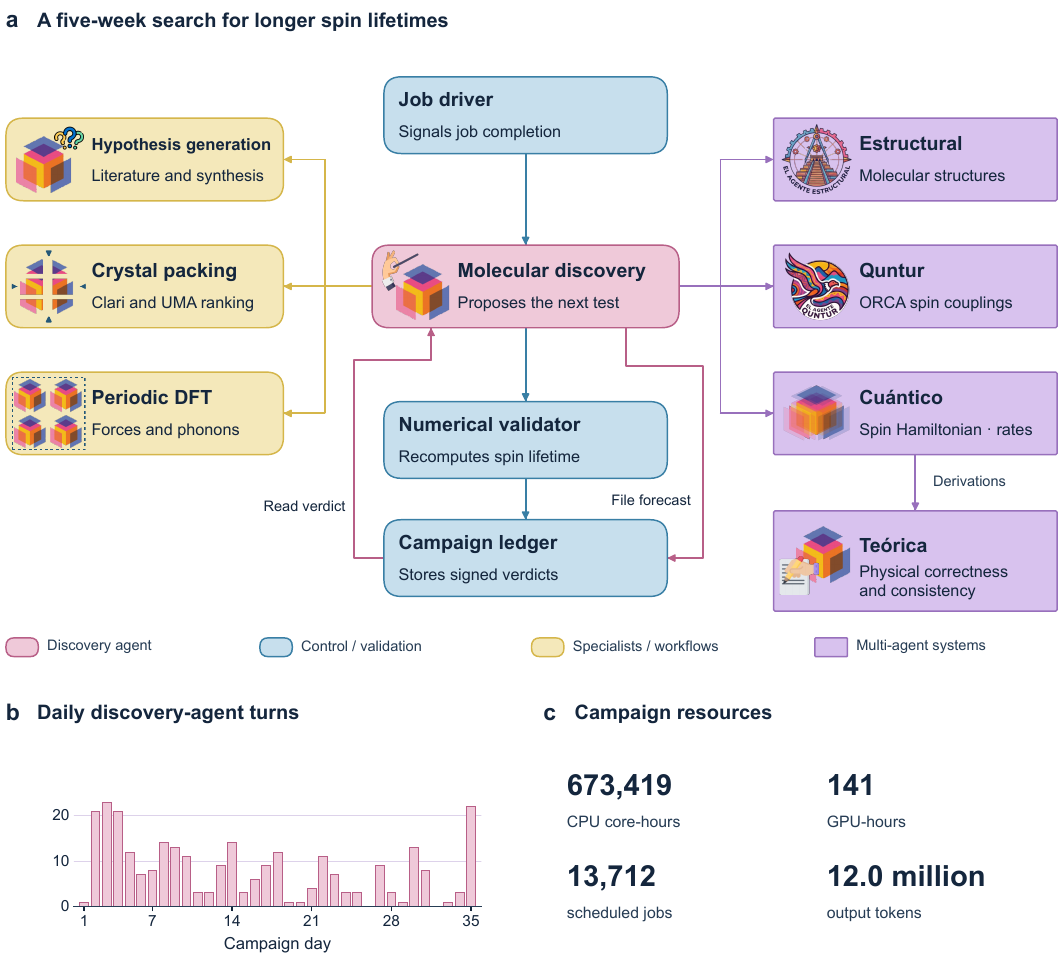}\par
{\raggedright\rmfamily\small \textbf{Fig. 1. The agent network and its five-week search for longer spin lifetimes.} a, The discovery agent proposed tests and the validator recomputed lifetimes. Quntur’s computational chemist coordinated a quantum chemistry specialist, an ORCA runner, an input-file service and a file-handling assistant. The 13 logged roles span the campaign. Cuántico sent its derivations to Teórica for checks. b, Daily discovery-agent turns. c, CPU time includes the pre-launch parent calculation. GPU time includes the parallel packing screen and UMA tests.\par}
\end{figure}
\subsection*{The parent vanadyl salt as a benchmark and starting structure}
We chose this vanadyl salt because its crystal structure is known and its long room-temperature spin lifetime has been measured experimentally\textsuperscript{\hyperlink{ref3}{3}}. The same compound therefore provides a structural starting point for substitutions and an experimental benchmark for the lifetime calculation. The parent salt, $(\mathrm{PPh}_4)_2[\mathrm{VO(dmit)}_2]$, contains two spin-bearing molecules per primitive cell, the smallest repeating unit of the crystal (Supplementary Fig. S1c). Each has two sulfur-rich ligands, 1,3-dithiole-2-thione-4,5-dithiolate (dmit). Each ligand can tilt about the line joining the two sulfur atoms that bind it to vanadium. We measure this fold relative to the plane through vanadium and those sulfurs. The larger parent fold is 15.4\(^{\circ}\) (Supplementary Fig. S1a,b). Each positive counterion has four carbon rings attached to phosphorus. These ligands and counterions provided two distinct sites for chemical substitution.

Before the search, our blind calculation gave a parent spin–lattice relaxation time of 1.76 \(\mu\)s, compared with 3.37 \(\mu\)s measured at 293 K for the magnetically diluted salt\textsuperscript{\hyperlink{ref3}{3}} (Supplementary Table S1). We also compared calculated temperature profiles with measurements on vanadyl phthalocyanine\textsuperscript{\hyperlink{ref1}{1}} and, in a blind test, copper octaethylporphyrin\textsuperscript{\hyperlink{ref34}{34}} (Supplementary Table S1). We used the calculated parent as the reference for every proposed salt, matching temperature, field and phonon sampling.\par
\par
The agents first considered vibrations that repeat in phase from cell to cell (zone-centre vibrations). At linewidth \(\sigma\) = 10 cm\(^{-1}\), one family accounted for 27.4\% of the calculated relaxation rate (Supplementary Fig. S1d). Vanadium moves against its four sulfur donors as the V=O bond tilts. At most 2.3\% of the vibrational weight lies on the counterion. The three ligand edits specified at launch tested whether changing these bonds would suppress this motion (\hyperlink{roundpath.2}{Fig. 2}). They replaced a thione sulfur with oxygen, rearranged the sulfur-rich ring or changed its composition (Supplementary Fig. S2b). The last structure also contained acetonitrile solvent. A fourth design tested the environment separately by replacing the neighbouring vanadyl anion with a diamagnetic molybdenyl analogue.
\par
Replacing the thione sulfur with oxygen weakened the targeted relaxation pathway, but other vibrations more than offset the gain and the spin relaxed faster overall. This result led the agents to keep the vanadyl molecule intact and change its neighbours instead (Fig. 2). Later calculations confirmed that all three ligand edits shortened the lifetime at linewidths of 7 cm\(^{-1}\) and above. In the molybdenyl host, the vanadyl molecule remained nearly flat and its lifetime exceeded the parent’s at narrow linewidth (Supplementary Tables S3 and S4).
\par
\subsection*{From ligand modifications to counterion substitutions}
\hyperlink{roundpath.2}{Figure 2} follows the search from the vanadyl ligands to the ions surrounding them. Once the ligand edits showed how readily other vibrations could offset a gain, the agents asked whether a neighbouring ion could restrain the same motions through its contacts with the vanadyl complex.
\par
\begin{figure}[!htbp]
\pdfbookmark[2]{Fig. 2. Results and subsequent hypotheses}{roundpath}
\centering\includegraphics[width=\linewidth]{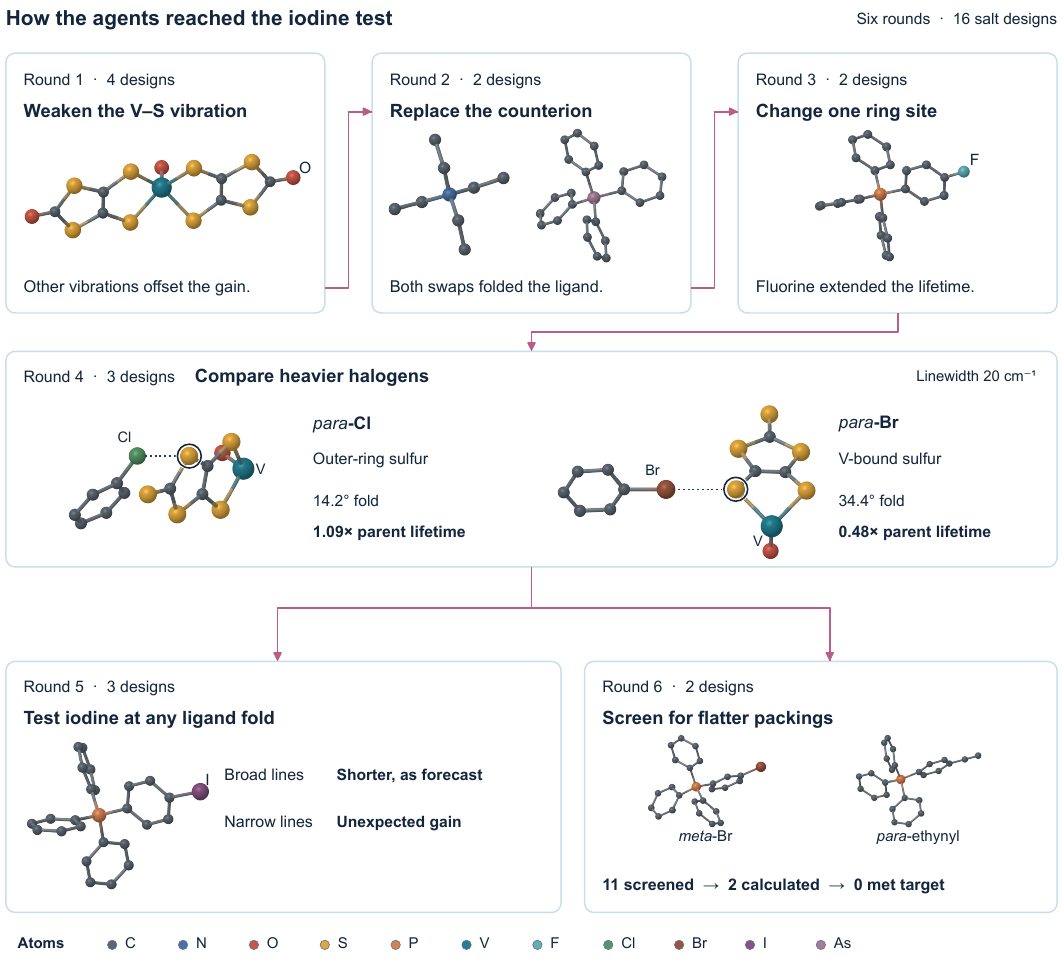}\par
{\raggedright\rmfamily\small \textbf{Fig. 2. How each round informed the next chemical test.} Boxes show representative designs and the number of salts taken to DFT. Pink arrows trace decisions. Circles mark the sulfur nearest chlorine or bromine. Lifetimes are relative to the parent at the zone centre and \(\sigma\) = 20 cm\(^{-1}\). The parallel branch screened for flatter packings. Atom colors: C, dark grey; N, blue; O, red; S, gold; P, orange; V, teal; F, turquoise; Cl, green; Br, brown; I, purple; As, mauve.\par}
\end{figure}
In round 2, the agents predicted crystal packings for two replacement counterions. Tetraethylammonium offered a way to constrain the sulfur atoms through close contacts. Tetraphenylarsonium preserved the four phenyl rings of the parent counterion while replacing phosphorus with arsenic. The arsenic substitution weakened the electronic response to the targeted vibrations, but in both predicted crystals a vanadyl ligand folded substantially and the calculated lifetime fell below the parent’s at broad linewidth. For the next round, the team kept the parent counterion and changed a single site on one of its rings (Fig. 2, rounds 2–3, and Supplementary Fig. S2a).
\par
At that site, opposite phosphorus on the ring, the agents tested two ideas. A cyano group might hold the vanadyl complex in place, while fluorine might alter the packing enough to weaken relaxation through other vibrations. In the zone-centre calculation at \(\sigma\) = 10 cm\(^{-1}\), the cyano salt relaxed faster than the parent and the fluorine salt slightly more slowly. The fluorine result warranted a follow-up because its predicted lifetime increased even though the vibrations expected to account for the gain contributed more strongly to relaxation. The team therefore varied the position of fluorine and the identity of the atom at the original site in separate tests (Fig. 2, rounds 3–4).
\par
\subsection*{Could a bend explain why the bromine compound relaxed faster?}
Moving fluorine one position around the ring removed its narrow-linewidth advantage. Keeping the original position and replacing fluorine with chlorine or bromine revealed a difference in how the counterion contacted the vanadyl complex (Fig. 3a,b). Chlorine lay closest to a sulfur on the outer ligand ring, whereas the carbon–bromine bond pointed almost directly at a sulfur bound to vanadium. Such directional contacts with sulfur occur in coordination cocrystals\textsuperscript{\hyperlink{ref35}{35}}. The ligand folded more in the bromine salt. At broad linewidth, its calculated spin lifetime fell below the parent’s, while the chlorine salt retained a slightly longer lifetime.
\par
\begin{figure}[!htbp]
\centering\includegraphics[width=\linewidth]{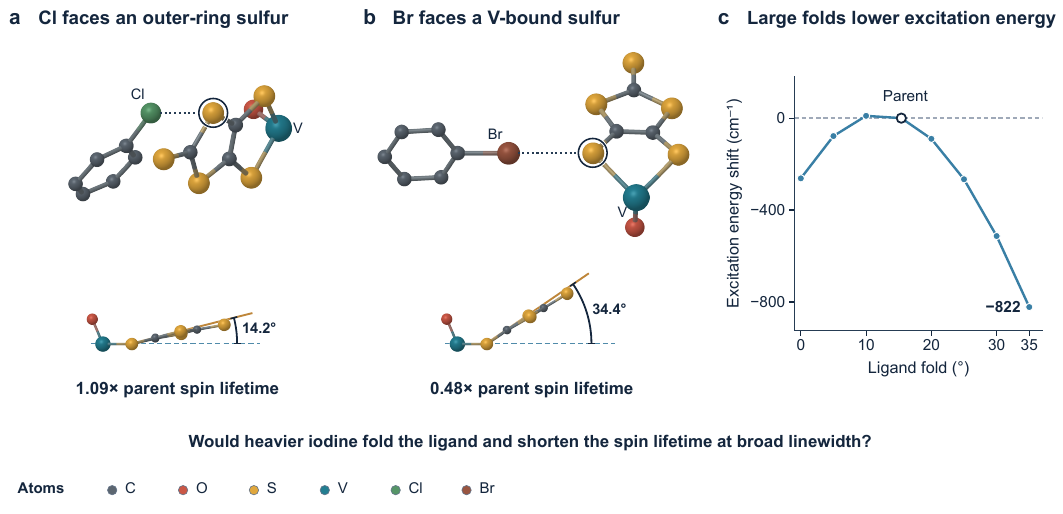}\par
{\raggedright\rmfamily\small \textbf{Fig. 3. Sulfur contacts and a controlled ligand bend motivate the iodine test.} a,b, Circles mark the nearest sulfur contacts in the chlorine salt (outer ring, 3.51 Å, 133\(^{\circ}\)) and bromine salt (vanadium-bound, 3.24 Å, 176\(^{\circ}\)). Angles are C–X···S. Lower views look along the sulfur hinge. Lifetime ratios use zone-centre vibrations at \(\sigma\) = 20 cm\(^{-1}\). c, Shift in the upper d\(\pi\) excitation as one parent ligand bends in a fixed crystal field. The open circle marks the parent fold (15.4\(^{\circ}\)). Atom colors: C, dark grey; O, red; S, gold; V, teal; Cl, green; Br, brown.\par}
\end{figure}
The chlorine and bromine structures differed in both ligand shape and the electric field around the vanadyl complex. To isolate the effect of shape, we bent one ligand in the parent structure while holding the surrounding field fixed (Fig. 3c). Studies of metal--dithiolate complexes show that folding can change the mixing of metal and sulfur orbitals\textsuperscript{\hyperlink{ref36}{36},\,\hyperlink{ref37}{37},\,\hyperlink{ref38}{38}}. At large folds, the upper vanadium d\(\pi\) excitation moved to lower energy, bringing an excited state involved in spin relaxation closer to the ground state. Separate calculations that exchanged the parent and chloro geometries and fields showed that both contributed to the higher excitation energy in the chlorine salt.
\par
The controlled bend established that folding could lower the excitation energy. Would a similar bend in another salt also shorten its spin lifetime?
\par
\subsection*{Iodine tested the prediction that a folded molecule would relax faster}
After the bromine result, the discovery agent chose iodine as the next halogen to test at the same ring position (Fig. 2, round 5). Before the packing calculation began, the discovery agent recorded two ways the predicted iodine structure could resemble the bromine salt. It could show a pronounced ligand fold or an iodine contact directed towards a sulfur bound to vanadium. The agent predicted that the iodine salt would have a shorter spin lifetime than the parent at broad phonon linewidth. Under the approved plan, the agent carried the iodine test through to a lifetime calculation regardless of which structure emerged.
\par
In the calculated iodine structure, one ligand folded beyond the forecast angle, although the nearest iodine–sulfur contact was with a sulfur on the outer ring (Fig. 4a,b). The zone-centre calculation also gave the predicted lifetime loss at broad linewidth (Fig. 4c). At the narrowest linewidth tested, the same folded structure had a longer spin lifetime than the parent. This narrow-linewidth gain was unexpected (Supplementary Fig. S3a,b). Supplementary Fig. S4a,b compares the structural forecast with the calculated packing.
\par
The chlorine result suggested a complementary test. The singly chlorinated counterion had left the ligands nearly flat, and the agent predicted that chlorinating additional rings would preserve that geometry. Both proposed salts instead exceeded the forecast fold angle (Fig. 4d). Under the agreed screening criteria, both stopped before the phonon calculation, leaving their predicted lifetimes untested (Supplementary Table S10).
\par
\begin{figure}[!htbp]
\centering\includegraphics[width=\linewidth]{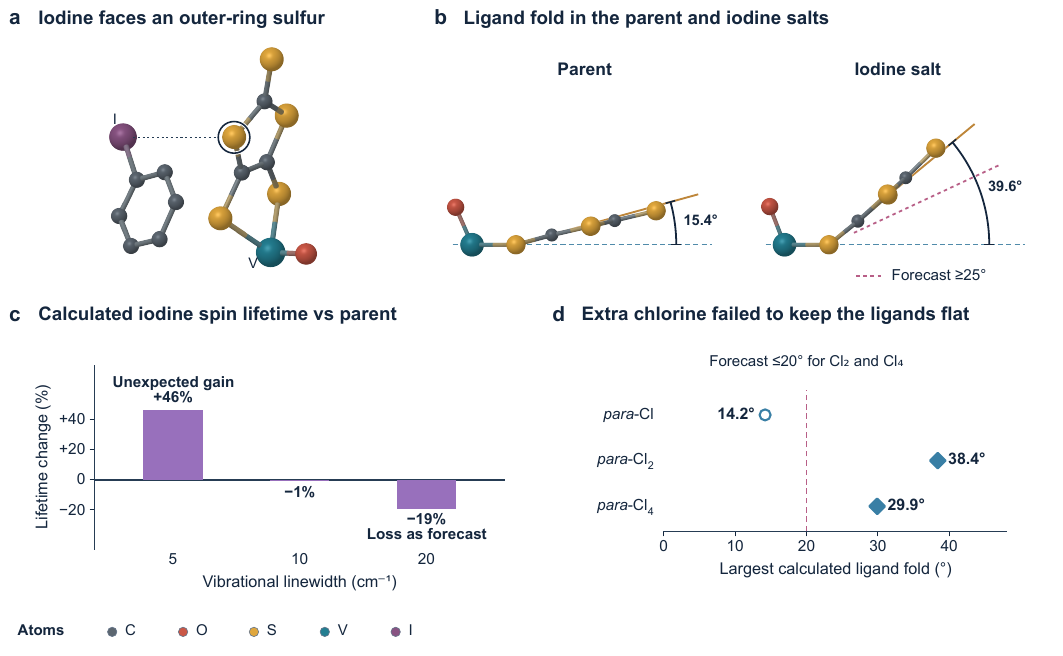}\par
{\raggedright\rmfamily\small \textbf{Fig. 4. Iodine lost spin lifetime as forecast at broad linewidth and gained it at narrow linewidth.} a, The nearest iodine contact is to an outer-ring sulfur (3.86 Å, 74\(^{\circ}\)). b, The 39.6\(^{\circ}\) fold exceeds the forecast threshold (dashed line). c, Iodine’s lifetime relative to the parent in the signed zone-centre calculation (300 K, 0.345 T, powder average). The loss at 20 cm\(^{-1}\) was forecast. The gain at 5 cm\(^{-1}\) was unexpected. d, Both additional chlorine substitutions exceeded the 20\(^{\circ}\) fold threshold and stopped before lifetime calculations. Atom colors: C, dark grey; O, red; S, gold; V, teal; I, purple.\par}
\end{figure}
\subsection*{A parallel screen selected two flatter packings}
The chlorine–bromine comparison suggested two tests of the same idea (Fig. 2). Iodine tested whether another strongly folded ligand would be associated with faster relaxation at broad linewidth. While that calculation was running, the agents pursued a second test. If a counterion kept the vanadyl ligands relatively flat, \textit{would the spin retain its energy for longer?}
\par
To search for such counterions, the agents compared the largest ligand fold in the lowest-energy packing predicted for each salt. After checking the UMA fold estimates against earlier DFT structures, they selected two counterions for refinement, one bearing bromine at the \textit{meta} position and the other an ethynyl group at the \textit{para} position (Supplementary Fig. S5a--c). Both refined structures folded slightly more than forecast, but their upper vanadium d\(\pi\) excitation energies met the criterion set in advance for proceeding to a lifetime calculation.
\par
In the zone-centre calculations, both salts had longer spin lifetimes than the parent at 10 cm\(^{-1}\) and shorter lifetimes at 20 cm\(^{-1}\) (Supplementary Table S10). The matched-grid comparison also placed both below the parent at 20 cm\(^{-1}\) (Supplementary Fig. S5d). Across the full set of models, every salt with a large ligand fold relaxed faster than the parent at the broader linewidth. The two screened salts showed the limit of that trend. Avoiding a large fold was insufficient to secure a longer lifetime at broad linewidth.
\par
\subsection*{Iodine and fluorine extend the predicted lifetime at narrow linewidth}
The parallel screen showed that selecting flatter ligands did not reliably prolong the spin lifetime. The lifetime calculations themselves had pointed to two candidates. Fluorine emerged from the first single-site substitutions, and iodine from the test of the bromine fold hypothesis. Both salts outlasted the parent at narrow linewidth in calculations that sampled only vibrations repeating in phase from one crystal cell to the next. Would that advantage survive when neighbouring cells could also vibrate out of phase?

We recalculated the lifetimes of all completed crystal models using each salt’s own DFT force constants and comparable sampling of vibrations across the crystal. At each linewidth we compared the candidate and parent under the same conditions (Fig. 5a).
\par
\begin{figure}[!htbp]
\centering\includegraphics[width=\linewidth]{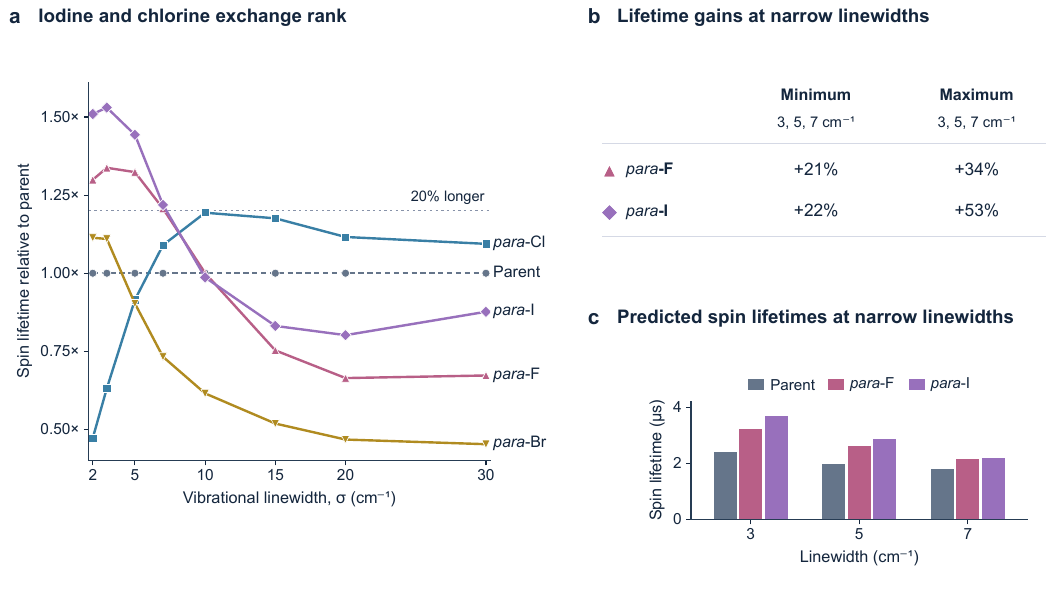}\par
{\raggedright\rmfamily\small \textbf{Fig. 5. Iodine leads the halogen series for narrow vibrational lines and chlorine for broad lines.} a, Lifetime relative to the parent at each phonon-pair broadening. b, Minimum and maximum gains at 3, 5 and 7 cm\(^{-1}\). c, Absolute lifetimes at those widths. Calculations use each salt’s DFT force constants, matched phonon grids and a powder average at 300 K and 0.345 T. Supplementary Tables S3–S4 cover all 17 models.\par}
\end{figure}

Both iodine and fluorine retained their advantage. At each of the three narrow linewidths compared in Fig. 5b, their calculated spin lifetimes exceeded the parent’s by more than 20\%. Figure 5c gives the absolute lifetimes of the parent and both candidates at 3, 5 and 7 cm\(^{-1}\). Iodine gave the longest lifetime in the halogen series at narrow linewidth, while chlorine led at broader linewidths (Fig. 5a). Supplementary Fig. S6a gives the lifetime curves for all crystal models, and Supplementary Fig. S6b compares each model with the parent at the same linewidth.
\par
Including vibrations beyond the zone centre did change the conclusion for another candidate. The tetraethylammonium salt lost its apparent narrow-linewidth gain under the same sampling check (Supplementary Fig. S7c). For iodine, the more complete calculation reproduced the reversal seen in the original test (Supplementary Fig. S4c). The spin relaxed more slowly than in the parent at narrow linewidth and faster at broad linewidth.\par
\par
\subsection*{Which pairs of vibrations change the ranking?}
\begin{figure}[!htbp]
\centering\includegraphics[width=\linewidth]{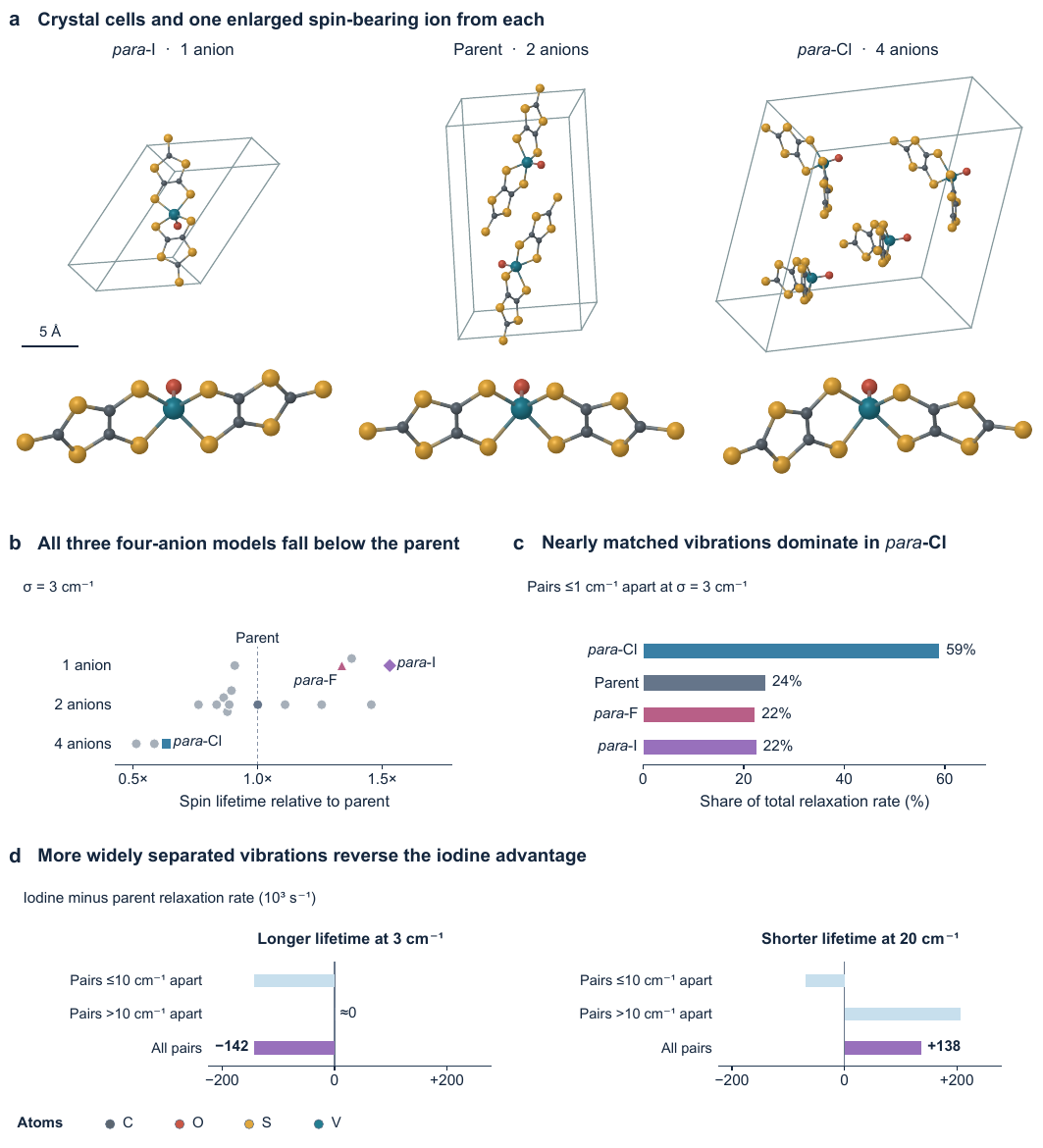}\par
{\raggedright\rmfamily\small \textbf{Fig. 6. Crystal packing and the vibration pairs that relax the spin.} a, Crystal cells at a common scale and aligned views of one anion per cell. b, Relative lifetimes at \(\sigma\) = 3 cm\(^{-1}\). Vertical offsets separate overlapping points. The dashed line marks the parent. c, Rate fraction from vibrations within 1 cm\(^{-1}\) of one another at \(\sigma\) = 3 cm\(^{-1}\). d, Iodine minus parent relaxation rates, partitioned at a frequency difference of 10 cm\(^{-1}\) (light blue) and summed (purple). Negative values indicate a longer iodine lifetime. Matched grids, 300 K and 0.345 T. Atom colors: C, dark grey; O, red; S, gold; V, teal.\par}
\end{figure}
Figure 5 showed that the same iodine structure could relax more slowly or more rapidly than the parent depending on the phonon linewidth. To trace that reversal, we grouped each pair of absorbed and emitted vibrations by the difference between their frequencies and compared their contributions to the relaxation rate (Fig. 6d). At narrow linewidth, nearly all relaxation in both salts came from pairs within 10 cm\(^{-1}\) of one another. These pairs made a smaller total contribution in iodine, giving it the longer spin lifetime.
\par
At broad linewidth, the contribution from those closer pairs was still smaller in iodine than in the parent. What changed was the contribution from vibrations farther apart in frequency. For example, a pair of iodine-salt modes near 210 and 228 cm\(^{-1}\) has little weight at narrow linewidth but can contribute at broad linewidth (Supplementary Note 2). Taken together, the more widely separated pairs added enough relaxation in iodine to outweigh its advantage from the closer pairs. This is where the lifetime ranking reversed (Fig. 6d).
\par
The packing comparison suggests one way that close-frequency pairs can arise. At narrow linewidth, all three calculation cells containing four spin-bearing anions gave shorter lifetimes than the parent (Fig. 6a,b). In the chlorine salt, two vanadyl anions undergo similar motions at nearly the same frequency, and the most closely spaced pairs account for a larger share of its relaxation than in iodine, fluorine or the parent (Fig. 6c). The relation between the number of anions and the lifetime weakened when approximately equivalent sites were counted together (Supplementary Table S6), so cell size alone does not establish the origin of these pairs.
\par
\pdfbookmark[1]{Discussion}{secDiscussion}
\section*{Discussion}
Changing a single site on a neighbouring ion led to two candidates for longer-lived vanadyl qubits. With narrow vibrational lines, the fluorine and iodine salts both had calculated room-temperature spin lifetimes more than 20\% longer than the parent’s, even when the calculation included vibrations moving out of phase in neighbouring cells (Fig. 5). The two candidates emerged for different reasons. Fluorine was chosen to slow relaxation. The discovery agent chose iodine to test whether the folding seen with bromine would again be accompanied by faster relaxation.
\par
The first ligand edit shows why it was useful to record both a forecast and the physical reason for it. Replacing sulfur with oxygen weakened the contribution from the vanadium–sulfur motion the agents had targeted, just as they proposed. Yet other vibrations carried away more energy, and the spin relaxed faster overall. That discrepancy gave the team a reason to keep the vanadyl molecule intact and change how its neighbours packed around it. Fluorine gave the converse result. Its calculated lifetime increased even though the vibrations the agents expected to suppress contributed more strongly to relaxation. Comparing individual pathways with the total rate let the team retain a promising candidate while revising the explanation for its performance.
\par
The chlorine and bromine tests gave the agents a more specific hypothesis. Chlorine left the ligand nearly flat, while bromine approached a sulfur bound to vanadium, bent the ligand and shortened the calculated lifetime at broad linewidth (Fig. 3). Five days before the iodine result, the agents forecast a similar lifetime loss. The calculation supported that forecast and overturned another from the same round. Adding chlorine to more rings failed to preserve the flat geometry of the singly chlorinated salt (Fig. 4). The parallel screen clarified how far the folding rule could be taken. A large fold consistently accompanied faster relaxation at broad linewidth in the salts studied, but selecting flatter packings alone did not secure a longer lifetime.
\par
The unexpected iodine result explains why the search had to consider competing pathways. At narrow linewidth, the spin exchanges energy mainly through pairs of vibrations close in frequency. Their combined contribution was smaller in iodine than in the parent. At broad linewidth, pairs farther apart in frequency contributed enough extra relaxation in iodine to reverse the ranking (Fig. 6d). The calculations resolve the contributions of individual pairs within each salt, although the correspondence between the parent’s targeted vibration and the iodine modes remains open. Chlorine provides a useful comparison. Similar motions on neighbouring vanadyl molecules form a close-frequency pair (Fig. 6c). Experiments that vary the spacing of spins or their surroundings\textsuperscript{\hyperlink{ref39}{39},\,\hyperlink{ref40}{40},\,\hyperlink{ref41}{41},\,\hyperlink{ref42}{42},\,\hyperlink{ref43}{43},\,\hyperlink{ref44}{44},\,\hyperlink{ref45}{45}} could help distinguish the effects of packing, molecular distortion\textsuperscript{\hyperlink{ref46}{46}}, spin dilution\textsuperscript{\hyperlink{ref47}{47}} and the way vibrations travel through the crystal\textsuperscript{\hyperlink{ref48}{48}}.
\par
The linewidths that favour iodine and fluorine are experimentally accessible in molecular crystals at room temperature. Our calculation assigns a width to a pair of vibrations. Assuming two equally broad Lorentzian peaks, the narrow range in Fig. 5 corresponds to individual peaks about 3.5–8.2 cm\(^{-1}\) wide at half their maximum height (Methods). At 300 K, Asher et al.\textsuperscript{\hyperlink{ref49}{49}} measured widths of about 4.6 cm\(^{-1}\) in ditBu-BTBT and 7 cm\(^{-1}\) in TIPS-pentacene (their Figs. 3b and 4b). These measurements give a concrete range to look for in the vanadyl crystals. Nabi et al. found that approximate phonon linewidths were sufficient to reproduce relaxation rates in a single-molecule magnet\textsuperscript{\hyperlink{ref50}{50}}.
\par
A direct experiment would compare the parent, fluorine, iodine and chlorine salts. Diffraction would establish whether the predicted packings and ligand folds form. Raman\textsuperscript{\hyperlink{ref34}{34}}, neutron\textsuperscript{\hyperlink{ref51}{51},\,\hyperlink{ref52}{52}} or far-infrared spectroscopy\textsuperscript{\hyperlink{ref53}{53}} could identify the vibrations and constrain their widths. Pulsed electron paramagnetic resonance can resolve the orientation dependence of spin relaxation\textsuperscript{\hyperlink{ref54}{54},\,\hyperlink{ref55}{55}} and probe molecular qubits in thin films\textsuperscript{\hyperlink{ref56}{56}}. For these salts, we would measure room-temperature lifetimes at matched field, orientation and spin dilution. Chlorine provides a test of the proposed mechanism. The calculations predict that iodine leads the halogen series when vibrational lines are narrow and chlorine leads when they are broader. Measuring both linewidths and spin lifetimes would test whether the crystals follow that predicted ordering.
\par
Using one result to choose the next test has precedents in reaction discovery\textsuperscript{\hyperlink{ref57}{57},\,\hyperlink{ref58}{58},\,\hyperlink{ref59}{59}}, reaction optimization\textsuperscript{\hyperlink{ref60}{60},\,\hyperlink{ref61}{61}}, materials searches\textsuperscript{\hyperlink{ref62}{62},\,\hyperlink{ref63}{63},\,\hyperlink{ref64}{64}}, solid-state synthesis\textsuperscript{\hyperlink{ref65}{65},\,\hyperlink{ref66}{66}} and diffraction\textsuperscript{\hyperlink{ref67}{67}}. In the vanadyl search, a salt chosen to test a predicted loss became a candidate for a longer-lived qubit. If the relevant vibrations have narrow lines at room temperature, we predict that both the iodine and fluorine salts will outlast the parent. Measurements of crystal packing, vibrational widths and spin lifetimes would give the agents an experimental basis for choosing the next counterion.
\par
The same pipeline could guide discovery wherever a chemical change improves one property at the expense of another. For a molecular quantum emitter, a substituent might shift emission to the desired wavelength while opening a route for the excited molecule to lose energy as heat. Specialists could identify the motion responsible and test whether stiffening that part of the molecule preserves both colour and brightness. Resonance Raman spectroscopy could also help distinguish molecular vibrations from solvent motion when optical coherence is the target\textsuperscript{\hyperlink{ref68}{68}}. For a battery electrolyte, a solvent that dissolves more lithium salt might also hold lithium ions too tightly for rapid transport. Simulations of how ions bind and move could guide the choice of a solvent mixture that balances salt solubility with ionic conductivity. For a porous framework that harvests water from dry air\textsuperscript{\hyperlink{ref19}{19}}, a water-binding group might improve uptake at low humidity while leaving less room in the pore. The agents could test whether moving the group closer to a pore wall preserves capacity. In each search, the team could record which experimental outcome would support its explanation before proposing the next molecule, just as the iodine forecast made an unexpected gain scientifically informative.
\par
\pdfbookmark[1]{Methods}{secMethods}
\section*{Methods}
Cuántico formulated the spin Hamiltonian and modeled how pairs of crystal vibrations relax the spin through excited electronic states. La Agente Teórica checked the algebra and notation of the derivations. Equation (1) quantifies how a single vibration changes the molecule's electronic states. Equation (2) combines the paths by which a pair of vibrations can change the spin state. Equation (3) weights each pair by its thermal occupation and by how closely its energy difference matches the energy exchanged with the spin.
\par
\subsection*{Search design and decision records}
Automated laboratories have linked chemical planning to synthesis and characterization\textsuperscript{\hyperlink{ref69}{69},\,\hyperlink{ref70}{70},\,\hyperlink{ref71}{71},\,\hyperlink{ref72}{72},\,\hyperlink{ref73}{73},\,\hyperlink{ref74}{74}}. Reviews describe how language-model agents organize such work\textsuperscript{\hyperlink{ref75}{75},\,\hyperlink{ref76}{76}}. Implemented systems select and execute tools\textsuperscript{\hyperlink{ref18}{18},\,\hyperlink{ref77}{77},\,\hyperlink{ref78}{78},\,\hyperlink{ref79}{79}}, assemble materials datasets from the literature\textsuperscript{\hyperlink{ref80}{80}}, coordinate chemistry research tasks\textsuperscript{\hyperlink{ref81}{81}} and evaluate candidate alloys\textsuperscript{\hyperlink{ref82}{82}}, inorganic crystals\textsuperscript{\hyperlink{ref83}{83}} and metal--organic frameworks\textsuperscript{\hyperlink{ref84}{84}}. Our campaign used specialist agents for the molecular and periodic calculations, following work on agent-assisted simulation protocols\textsuperscript{\hyperlink{ref85}{85},\,\hyperlink{ref86}{86},\,\hyperlink{ref87}{87},\,\hyperlink{ref88}{88},\,\hyperlink{ref89}{89},\,\hyperlink{ref90}{90}}. A persistent record connected each calculation to the hypothesis that prompted it.
\par
The molecular-discovery agent (gpt-6-astra, high reasoning effort) read a persistent campaign ledger before proposing each new test. The ledger contained the reference structure, accepted scope, chemical changes, hypotheses, job identifiers and verified outcomes. Specialist agents (Claude Opus 5.5) handled hypothesis formulation, literature mining with Undermind, synthesis and geometry checks, periodic DFT, quantum chemistry and quantum physics. Some specialists delegated further. Within Quntur, a quantum-chemistry specialist called the ORCA runner. El Agente Cuántico sent its derivations to La Agente Teórica for symbolic checks. Tool-running agents used Claude Sonnet 5. A scheduler driver resumed the discovery agent when a job finished or a candidate had no work running. Of the 1,365 agent invocations, 1,279 resumed from a condensed record of earlier work. The thread logs record 18,028 tool calls, while the scheduler counted 13,712 jobs. CPU use was computed as allocated cores times elapsed hours. Supplementary Fig. S8a shows when each specialist was active, and Supplementary Fig. S8b gives the number of calls to each scientific tool. Supplementary Fig. S3c breaks down the CPU time for the individual salts by calculation stage. The reported GPU time covers the scientific calculations.
\par
The principal investigator specified the objective and approved extensions and screening criteria. The discovery agent chose iodine as the round-5 test of the trend seen with bromine. The investigator approved the round, including the stopping criteria for the two chlorinated salts and a full lifetime calculation for iodine regardless of its fold. The discovery agent recorded the forecasts against the same parent structure before submitting the packing calculations. The original forecast, ledger export and signed iodine receipt accompany the source data (Supplementary Table S10). A supervising AI coordinator recorded those decisions, handled large displacement arrays and memory requirements beyond the agents' scheduling tools, and carried out the comparisons across salts, including the fold scan and matched-grid linewidth analysis. Following established practice for electronic-structure workflows\textsuperscript{\hyperlink{ref91}{91},\,\hyperlink{ref92}{92}}, the record linked each result to the inputs, settings and jobs that produced it. A separate language-model judge checked whether the evidence supported each claimed campaign milestone.
\par
Only the numerical validator could enter a lifetime in the campaign ledger. It read the submitted phonons, displacement manifest and vibronic couplings, then ran the same rate routine used in the campaign to recompute $T_1$. The iodine receipt includes coupling sets at two displacement amplitudes. The validator also aligned the donor coordination shells and compared the vibrational subspaces of the parent and candidate to test whether the target motion persisted. For iodine the match failed. The receipt therefore records the total lifetime ratio and leaves its microscopic origin unassigned. The coordinator later compared lifetimes on matched phonon grids.
\par
\subsection*{Crystal packings and normal modes}
We used reported crystal structures of the parent\textsuperscript{\hyperlink{ref3}{3}}, the carbonyl and sulfur-connectivity variants\textsuperscript{\hyperlink{ref5}{5}}, and the C$_5$H$_6$S$_4$ complex\textsuperscript{\hyperlink{ref93}{93}} (Supplementary Table S5). We retained alternative ordered models within 2 kJ mol\(^{-1}\) per formula unit. For the host control we replaced one molybdenyl centre in the isostructural host with vanadyl. For salts without a deposited structure, CLARI-H\textsuperscript{\hyperlink{ref32}{32}} used a standard request of 100 trial packings for each cell size (one, two or four formula units). We relaxed the packings with the molecular-crystal task of UMA-s-1p2\textsuperscript{\hyperlink{ref33}{33}} to a maximum force of 0.05 eV \AA{}\(^{-1}\). The packing search used 300 K free-energy estimates and sent up to 15 candidates within 10 kJ mol\(^{-1}\) per formula unit to DFT.
\par
We ranked the packings by spin-unrestricted DFT in CP2K 2024.2\textsuperscript{\hyperlink{ref94}{94}} using PBE\textsuperscript{\hyperlink{ref95}{95}}, zero-damping D3 dispersion\textsuperscript{\hyperlink{ref96}{96}}, DZVP-MOLOPT-SR-GTH basis sets\textsuperscript{\hyperlink{ref97}{97}} and GTH pseudopotentials\textsuperscript{\hyperlink{ref98}{98},\,\hyperlink{ref99}{99}}. We used ferromagnetic alignment without a Hubbard correction, plane-wave and relative cutoffs of 800 and 60 Ry, and $\lceil20.6\,\text{Å}/L_j\rceil$ electronic $k$ points along each lattice vector of length $L_j$. We optimized the cell and then the ions at fixed cell, with a maximum force of $10^{-4}$ hartree per bohr and an electronic convergence threshold of $10^{-8}$ hartree.
\par
For the lowest-energy packing of each predicted salt, a one-sided Cartesian displacement of 0.01 \AA{} supplied one column of the force-constant matrix from the change in forces on all atoms. We used the same electronic Hamiltonian for these displacements and enlarged the cell if any width was below 9 \AA{}. For imaginary modes, we compared the energies at opposite displacements with the equilibrium energy. An increase in both directions supported a numerical origin. Diagonalizing the mass-weighted dynamical matrix gave the angular frequency $\omega_\lambda$ and displacement pattern of each phonon $\lambda$\textsuperscript{\hyperlink{ref100}{100}}. The frequency sets its energy $\epsilon_\lambda=\hbar\omega_\lambda$. The motion it induces on the vanadyl complex determines its coupling to the spin.
\par
\subsection*{Electronic response to a crystal vibration}
First-principles treatments of spin relaxation resolve the contributions of competing vibrational and electronic pathways\textsuperscript{\hyperlink{ref101}{101},\,\hyperlink{ref102}{102},\,\hyperlink{ref103}{103},\,\hyperlink{ref104}{104}}. Their rates can depend on electron correlation beyond an active space\textsuperscript{\hyperlink{ref105}{105}} and on the treatment of the crystal environment\textsuperscript{\hyperlink{ref50}{50},\,\hyperlink{ref106}{106}}. Machine-learning models can accelerate the evaluation of spin--phonon couplings\textsuperscript{\hyperlink{ref107}{107}}. We retained a multireference description of the vanadyl electronic states inside the electrostatic field of each packing. Iodine entered the crystal structure and phonon calculations. Its representation as embedding charges in the electronic calculation omitted direct spin--orbit coupling from the iodine atom.
\par
We extracted the vanadyl complex in the coordinate frame of the phonons and fitted 600 embedding charges to the Ewald potential of the formal crystal charges (10 \AA{} explicit region, 22 \AA{} fit radius). Quntur\textsuperscript{\hyperlink{ref31}{31}} autonomously selected an active space of one electron in five vanadium $d$ orbitals for the vanadyl series. We used this active space in ORCA 6.0\textsuperscript{\hyperlink{ref108}{108}} for CASSCF calculations averaged over five electronic states\textsuperscript{\hyperlink{ref109}{109}}. We applied strongly contracted second-order $N$-electron valence state perturbation theory\textsuperscript{\hyperlink{ref110}{110}} and RI-SOMF(1X) spin--orbit coupling\textsuperscript{\hyperlink{ref111}{111}}, using def2-TZVP for V, O and S and def2-SVP for the other atoms\textsuperscript{\hyperlink{ref112}{112}}. The five orbital states and two spin projections give a ground Kramers doublet and four excited doublets. Ab initio ligand-field analysis\textsuperscript{\hyperlink{ref113}{113}} placed the orbitals in a common metal frame with the $z$ axis along V=O.
\par
We varied the embedded geometry along mass-weighted directions spanning the local displacements of the zone-centre modes up to 500 cm\(^{-1}\). The embedding charges stayed at their equilibrium positions throughout the displacement sweep. Opposite displacements, typically 0.01 \AA{} in Cartesian space, gave the derivative of the electronic Hamiltonian in each direction. We checked linearity with doubled displacements and extended the vibrational range to 800 cm\(^{-1}\) when the cutoff test changed the lifetime by more than 5\%. For a real, mass-normalized mode coordinate $Q_\lambda$, the one-phonon coupling between equilibrium electronic states $r$ and $s$ is
\par
\begin{equation}
g^\lambda_{rs}=\sqrt{\frac{\hbar}{2\omega_\lambda}}\,
\left\langle r\left|\frac{\partial H_{\rm el}}{\partial Q_\lambda}\right|s\right\rangle .
\label{eq:spin1}
\end{equation}
The square root is the zero-point amplitude. Thus $g^\lambda_{rs}$ has units of energy and includes both how much the atoms move and how strongly that motion mixes the electronic states.
\par
\subsection*{Two-phonon spin relaxation}
Chemical changes that alter relaxation at intermediate temperatures can have little effect near room temperature\textsuperscript{\hyperlink{ref114}{114}}. We considered the Raman mechanism, in which the spin absorbs one phonon and emits another through a virtual electronic state\textsuperscript{\hyperlink{ref115}{115},\,\hyperlink{ref116}{116},\,\hyperlink{ref117}{117}}. Relaxation through a thermally populated state is a distinct mechanism\textsuperscript{\hyperlink{ref118}{118}}. The Raman rate depends on the energy mismatch between the phonon pair and the spin transition\textsuperscript{\hyperlink{ref119}{119},\,\hyperlink{ref120}{120}}, the vibrational linewidths\textsuperscript{\hyperlink{ref50}{50}} and the electronic states\textsuperscript{\hyperlink{ref121}{121}} whose energies depend on ligand geometry\textsuperscript{\hyperlink{ref122}{122},\,\hyperlink{ref123}{123}}. In our vanadyl phthalocyanine benchmark at 300 K, modulation of the ground-doublet $g$ and hyperfine tensors alone gave a lifetime about four orders of magnitude longer than measured. Virtual transitions through the excited $d$ states gave 0.84 \(\mu\)s, compared with the experimental 1.1 \(\mu\)s\textsuperscript{\hyperlink{ref1}{1}} (Supplementary Table S1). We therefore used the excited-state Raman channel to rank the salts.
\par
For an effective spin $1/2$, the electronic Zeeman term is $H_Z=\mu_{\rm B}\,\mathbf B\cdot\mathbf g\cdot\mathbf S$, where $\mathbf S$ is the dimensionless spin operator and $\mathbf g$ is the electronic $g$ tensor\textsuperscript{\hyperlink{ref121}{121}}. The two energies differ by $\mu_{\rm B}|\mathbf g^{\mathrm T}\mathbf B|$, so the splitting depends on the field direction. Consider a transition from component $i$ to $f$ at a fixed orientation. A phonon $\alpha$ is absorbed and a distinct phonon $\beta$ is emitted. In the excited-state Raman channel $\Gamma_{\rm II}$\textsuperscript{\hyperlink{ref121}{121}}, either event can occur first. In the sum-over-states description, the second-order amplitude for this transition is
\par
\begin{equation}
A^{\alpha\beta}_{fi}=\sum_{m\in\mathcal E}
\left[
\frac{g^\beta_{fm}g^\alpha_{mi}}{E_i+\epsilon_\alpha-E_m}
+
\frac{g^\alpha_{fm}g^\beta_{mi}}{E_i-\epsilon_\beta-E_m}
\right] .
\label{eq:spin2}
\end{equation}
Here $E_r$ is the energy of electronic state $r$, and $\mathcal E$ contains both components of the four excited doublets. The intermediate state is virtual. Its energy enters the denominator without a thermal population factor. The sum retains interference between electronic pathways. Its field-independent spin-changing component is related to the non-Abelian Berry curvature of the ground doublet\textsuperscript{\hyperlink{ref124}{124}}.
\par
With $n_\lambda=[\exp(\epsilon_\lambda/k_{\rm B}T)-1]^{-1}$ and $\Delta_{fi}=E_f-E_i$, the rate for this direction of spin transfer is
\par
\begin{equation}
W_{f\leftarrow i}=\frac{2\pi}{\hbar}\sum_{\alpha\ne\beta}
|A^{\alpha\beta}_{fi}|^2 n_\alpha(n_\beta+1)\,
\frac{\exp[-(\epsilon_\alpha-\epsilon_\beta-\Delta_{fi})^2/(2\eta^2)]}{\sqrt{2\pi}\,\eta},
\qquad \eta=hc\sigma .
\label{eq:spin3}
\end{equation}
The sum runs over distinct retained modes with positive frequency. The rate is fourth order in the linear couplings $g$ because it contains the squared second-order amplitude. Here $T$ is the temperature, $k_{\rm B}$ is Boltzmann's constant, $h$ is Planck's constant and $c$ is the speed of light, expressed in centimetres per second when $\sigma$ is in cm\(^{-1}\). The Gaussian is normalized in energy. The reported $\sigma$ is its standard deviation in cm\(^{-1}\). At 0.345 T the spin splitting is about 0.3 cm\(^{-1}\), so narrow lines select phonons of nearly equal frequency. Broadening the lines admits pairs with larger mismatches, whose couplings can be quite different. For the two ground-state populations, the fixed-orientation longitudinal rate is $T_1^{-1}=W_{f\leftarrow i}+W_{i\leftarrow f}$. The salt comparisons used 300 K and 0.345 T. The powder-averaged rate was obtained as the weighted mean of $1/T_1$ over a 14-point Lebedev grid of field orientations\textsuperscript{\hyperlink{ref125}{125}} (6 points with weight 1/15, 8 with weight 3/40), and the reported $T_1$ is the inverse of this mean rate.
\par
\subsection*{Reciprocal-space sampling and comparison criteria}
During the search, the agents used zone-centre phonons to diagnose vibrational families and evaluate their forecasts. For the final rankings we evaluated each salt's own DFT force constants on a grid whose equivalent supercell was at least 21 \AA{} wide in every direction. We projected the resulting normal modes onto that salt's computed local electronic derivatives (Supplementary Note 1). The parent, \textit{para}-I and \textit{para}-F grids gave 1,552, 1,237 and 1,816 included modes, respectively. Across the 42 zone-centre lifetimes tabulated in both the validator records and the reconstruction, the largest difference was 0.46\% (\textit{para}-Br at 5 cm\(^{-1}\)). Supplementary Fig. S7a compares the parent lifetime with zone-centre and denser phonon sampling. Recalculating the parent force constants in a $2\times1\times1$ DFT supercell changed its matched-grid lifetime by less than 0.4\% at every sampled width (Supplementary Fig. S7b and Table S2).
\par
At each of eight widths between 2 and 30 cm\(^{-1}\) we divided the salt's powder lifetime by that of the parent under the same sampling rule. Figure 5b reports the range of gains across 3, 5 and 7 cm\(^{-1}\). For two Lorentzian phonons of equal full width at half maximum $w$, the pair line has width $2w$. Matching this width to the Gaussian gives $2w=2\sqrt{2\ln 2}\,\sigma$. The three narrow broadenings correspond to individual widths of about 3.5, 6 and 8 cm\(^{-1}\) under that convention. The mode linewidths of the proposed salts remain to be measured.
\par
Spearman coefficients treat the 17 crystal models as observations, including each ordered structure where alternatives were available. For the correlation with the rate share from closely spaced phonon pairs, all 17 models had dense-grid lifetimes and zone-centre pair decompositions. The four salts in Fig. 6c also had dense-grid pair decompositions. These fractions measure how much each pair contributes to spin relaxation. A complementary approach tracks the energy deposited in each part of the environment when its motion drives transitions between electronic states\textsuperscript{\hyperlink{ref126}{126}}. Supplementary Note 2 defines the pair fractions and shows how electronic pathways interfere within each pair.
\par
\subsection*{Fold scan and molecular graphics}
For each ligand we measured the acute angle between the plane containing vanadium and its two donor sulfurs and the least-squares plane through those sulfurs and their two adjacent carbons. For each packing we report the larger of the two folds on the vanadyl anion used for the electronic calculation. In the controlled scan we changed one parent fold from 0\(^{\circ}\) to 35\(^{\circ}\) in the fixed parent embedding and recalculated the upper $d\pi$ energy.
\par
The molecular graphics are orthographic projections of the calculated coordinates. Atom colors throughout the molecular drawings: C, dark grey; H, light grey; N, blue; O, red; S, gold; P, orange; V, teal; F, turquoise; Cl, green; Br, brown; I, purple; As, mauve. Each figure key lists only the elements shown. The fold profiles look along the donor sulfur pair with the VS\textsubscript{2} plane horizontal. We identified the nearest periodic halogen--sulfur pair and measured the C--X\(\cdots\)S angle in the plane of the three atoms. Counterion fragments share a common face-on ring orientation. We reconstructed whole ions across periodic boundaries and omitted carbon-bound hydrogens except where one marks the parent substitution site. We used Python and Matplotlib for the structures and quantitative plots.
\par
\pdfbookmark[1]{Data availability}{secDataavailability}
\section*{Data availability}
The dataset contains relaxed crystal structures, phonon force constants, electronic couplings, displacement manifests and embedding charges for the parent vanadyl salt and 16 related crystal models. It also includes the calculated lifetimes at each linewidth, the data underlying the figures and the campaign records. The data are available from \href{https://doi.org/10.5683/SP4/SG6QHU}{Borealis (10.5683/SP4/SG6QHU)}. The interactive \href{https://igustin-nv.github.io/molecular-qubit-atlas/}{Molecular Qubit Atlas} lets readers rotate crystal structures, animate vibrational modes and explore the data behind the figures.
\par
\pdfbookmark[1]{Code availability}{secCodeavailability}
\section*{Code availability}
The Python code deposited on \href{https://doi.org/10.5683/SP4/SG6QHU}{Borealis (10.5683/SP4/SG6QHU)} recomputes room-temperature spin lifetimes from the phonon and electronic inputs. The package includes a worked example for the parent and iodine salts, a test against the reported lifetimes and scripts to regenerate the article and supplementary figures.
\par
\pdfbookmark[1]{Acknowledgements}{secAcknowledgements}
\section*{Acknowledgements}

We gratefully acknowledge the longstanding contributions of current and former members of the \href{https://matter.toronto.edu/}{Matter Lab}, particularly the El Agente team. I.G. and J.B.P.S. acknowledge funding for this project from the Natural Sciences and Engineering Research Council of Canada (NSERC) through Alliance Grant ALLRP587593-23.
\par
A.A.-G. thanks Anders G. Frøseth for his generous support. A.A.-G. and V.B. gratefully acknowledge support from Natural Resources Canada, the Canada 150 Research Chairs program and the University of Toronto’s Acceleration Consortium. The Acceleration Consortium receives funding from the Canada First Research Excellence Fund (CFREF-2022-00042).
This work was supported by the Defense Advanced Research Projects Agency (DARPA) under Agreement No. HR0011262E022 and by the AI2050 program of Schmidt Sciences.

This research was enabled in part by support provided by the SciNet HPC Consortium (\href{https://scinethpc.ca}{scinethpc.ca}), Calcul Qu\'ebec (\href{https://www.calculquebec.ca}{www.calculquebec.ca}), and the Digital Research Alliance of Canada (\href{https://alliancecan.ca}{alliancecan.ca}).
\par

\pdfbookmark[1]{References}{secReferences}
\section*{References}
{\fontsize{8.6}{9.65}\selectfont\setlength{\parskip}{0.12pt}
\hypertarget{ref1}{}\hangindent=2.45em\hangafter=1\noindent\makebox[2.45em][l]{1.}Atzori, M. et al. Room-Temperature Quantum Coherence and Rabi Oscillations in Vanadyl Phthalocyanine: Toward Multifunctional Molecular Spin Qubits. \emph{Journal of the American Chemical Society} \textbf{138}, 2154--2157 (2016). \href{https://doi.org/10.1021/jacs.5b13408}{DOI}
\par
\hypertarget{ref2}{}\hangindent=2.45em\hangafter=1\noindent\makebox[2.45em][l]{2.}Bayliss, S. L. et al. Optically addressable molecular spins for quantum information processing. \emph{Science} \textbf{370}, 1309--1312 (2020). \href{https://doi.org/10.1126/science.abb9352}{DOI}
\par
\hypertarget{ref3}{}\hangindent=2.45em\hangafter=1\noindent\makebox[2.45em][l]{3.}Atzori, M. et al. Quantum Coherence Times Enhancement in Vanadium(IV)-based Potential Molecular Qubits: the Key Role of the Vanadyl Moiety. \emph{Journal of the American Chemical Society} \textbf{138}, 11234--11244 (2016). \href{https://doi.org/10.1021/jacs.6b05574}{DOI}
\par
\hypertarget{ref4}{}\hangindent=2.45em\hangafter=1\noindent\makebox[2.45em][l]{4.}Du, J.-L., Eaton, G. R. \& Eaton, S. S. Electron Spin Relaxation in Vanadyl, Copper(II), and Silver(II) Porphyrins in Glassy Solvents and Doped Solids. \emph{Journal of Magnetic Resonance, Series A} \textbf{119}, 240--246 (1996). \href{https://doi.org/10.1006/jmra.1996.0079}{DOI}
\par
\hypertarget{ref5}{}\hangindent=2.45em\hangafter=1\noindent\makebox[2.45em][l]{5.}Yu, C.-J. et al. Long Coherence Times in Nuclear Spin-Free Vanadyl Qubits. \emph{Journal of the American Chemical Society} \textbf{138}, 14678--14685 (2016). \href{https://doi.org/10.1021/jacs.6b08467}{DOI}
\par
\hypertarget{ref6}{}\hangindent=2.45em\hangafter=1\noindent\makebox[2.45em][l]{6.}Zadrozny, J. M., Niklas, J., Poluektov, O. G. \& Freedman, D. E. Millisecond Coherence Time in a Tunable Molecular Electronic Spin Qubit. \emph{ACS Central Science} \textbf{1}, 488--492 (2015). \href{https://doi.org/10.1021/acscentsci.5b00338}{DOI}
\par
\hypertarget{ref7}{}\hangindent=2.45em\hangafter=1\noindent\makebox[2.45em][l]{7.}Graham, M. J. et al. Influence of Electronic Spin and Spin--Orbit Coupling on Decoherence in Mononuclear Transition Metal Complexes. \emph{Journal of the American Chemical Society} \textbf{136}, 7623--7626 (2014). \href{https://doi.org/10.1021/ja5037397}{DOI}
\par
\hypertarget{ref8}{}\hangindent=2.45em\hangafter=1\noindent\makebox[2.45em][l]{8.}Jackson, C. E. et al. Impact of Counter Ion Methyl Groups on Spin Relaxation in [V(C6H4O2)3]2--. \emph{The Journal of Physical Chemistry C} \textbf{126}, 7169--7176 (2022). \href{https://doi.org/10.1021/acs.jpcc.2c01090}{DOI}
\par
\hypertarget{ref9}{}\hangindent=2.45em\hangafter=1\noindent\makebox[2.45em][l]{9.}Gustin, I., Chen, X. \& Franco, I. Decoherence dynamics in molecular qubits: Exponential, Gaussian and beyond. \emph{The Journal of Chemical Physics} \textbf{162}, 064106 (2025). \href{https://doi.org/10.1063/5.0246970}{DOI}
\par
\hypertarget{ref10}{}\hangindent=2.45em\hangafter=1\noindent\makebox[2.45em][l]{10.}Tesi, L. et al. Quantum coherence in a processable vanadyl complex: new tools for the search of molecular spin qubits. \emph{Chemical Science} \textbf{7}, 2074--2083 (2016). \href{https://doi.org/10.1039/c5sc04295j}{DOI}
\par
\hypertarget{ref11}{}\hangindent=2.45em\hangafter=1\noindent\makebox[2.45em][l]{11.}Atzori, M. et al. Spin Dynamics and Low Energy Vibrations: Insights from Vanadyl-Based Potential Molecular Qubits. \emph{Journal of the American Chemical Society} \textbf{139}, 4338--4341 (2017). \href{https://doi.org/10.1021/jacs.7b01266}{DOI}
\par
\hypertarget{ref12}{}\hangindent=2.45em\hangafter=1\noindent\makebox[2.45em][l]{12.}Atzori, M. et al. Structural Effects on the Spin Dynamics of Potential Molecular Qubits. \emph{Inorganic Chemistry} \textbf{57}, 731--740 (2018). \href{https://doi.org/10.1021/acs.inorgchem.7b02616}{DOI}
\par
\hypertarget{ref13}{}\hangindent=2.45em\hangafter=1\noindent\makebox[2.45em][l]{13.}Follmer, A. H., Ribson, R. D., Oyala, P. H., Chen, G. Y. \& Hadt, R. G. Understanding Covalent versus Spin--Orbit Coupling Contributions to Temperature-Dependent Electron Spin Relaxation in Cupric and Vanadyl Phthalocyanines. \emph{The Journal of Physical Chemistry A} \textbf{124}, 9252--9260 (2020). \href{https://doi.org/10.1021/acs.jpca.0c07860}{DOI}
\par
\hypertarget{ref14}{}\hangindent=2.45em\hangafter=1\noindent\makebox[2.45em][l]{14.}Fataftah, M. S. et al. Metal--ligand covalency enables room temperature molecular qubit candidates. \emph{Chemical Science} \textbf{10}, 6707--6714 (2019). \href{https://doi.org/10.1039/c9sc00074g}{DOI}
\par
\hypertarget{ref15}{}\hangindent=2.45em\hangafter=1\noindent\makebox[2.45em][l]{15.}Kazmierczak, N. P., Mirzoyan, R. \& Hadt, R. G. The Impact of Ligand Field Symmetry on Molecular Qubit Coherence. \emph{Journal of the American Chemical Society} \textbf{143}, 17305--17315 (2021). \href{https://doi.org/10.1021/jacs.1c04605}{DOI}
\par
\hypertarget{ref16}{}\hangindent=2.45em\hangafter=1\noindent\makebox[2.45em][l]{16.}Amdur, M. J. et al. Chemical control of spin--lattice relaxation to discover a room temperature molecular qubit. \emph{Chemical Science} \textbf{13}, 7034--7045 (2022). \href{https://doi.org/10.1039/d1sc06130e}{DOI}
\par
\hypertarget{ref17}{}\hangindent=2.45em\hangafter=1\noindent\makebox[2.45em][l]{17.}Angello, N. H. et al. Closed-loop transfer enables artificial intelligence to yield chemical knowledge. \emph{Nature} \textbf{633}, 351--358 (2024). \href{https://doi.org/10.1038/s41586-024-07892-1}{DOI}
\par
\hypertarget{ref18}{}\hangindent=2.45em\hangafter=1\noindent\makebox[2.45em][l]{18.}Boiko, D. A., MacKnight, R., Kline, B. \& Gomes, G. Autonomous chemical research with large language models. \emph{Nature} \textbf{624}, 570--578 (2023). \href{https://doi.org/10.1038/s41586-023-06792-0}{DOI}
\par
\hypertarget{ref19}{}\hangindent=2.45em\hangafter=1\noindent\makebox[2.45em][l]{19.}Jaffrelot Inizan, T. et al. System of agentic AI for the discovery of metal--organic frameworks. \emph{Nature Computational Science} \textbf{6}, 1223--1234 (2026). \href{https://doi.org/10.1038/s43588-026-01058-x}{DOI}
\par
\hypertarget{ref20}{}\hangindent=2.45em\hangafter=1\noindent\makebox[2.45em][l]{20.}Gottweis, J. et al. Accelerating scientific discovery with Co-Scientist. \emph{Nature} \textbf{655}, 487--496 (2026). \href{https://doi.org/10.1038/s41586-026-10644-y}{DOI}
\par
\hypertarget{ref21}{}\hangindent=2.45em\hangafter=1\noindent\makebox[2.45em][l]{21.}Ghareeb, A. E. et al. A multi-agent system for automating scientific discovery. \emph{Nature} \textbf{655}, 497--505 (2026). \href{https://doi.org/10.1038/s41586-026-10652-y}{DOI}
\par
\hypertarget{ref22}{}\hangindent=2.45em\hangafter=1\noindent\makebox[2.45em][l]{22.}Swanson, K. et al. The Virtual Lab of AI agents designs new SARS-CoV-2 nanobodies. \emph{Nature} \textbf{646}, 716--723 (2025). \href{https://doi.org/10.1038/s41586-025-09442-9}{DOI}
\par
\hypertarget{ref23}{}\hangindent=2.45em\hangafter=1\noindent\makebox[2.45em][l]{23.}Zou, Y. et al. El Agente: An autonomous agent for quantum chemistry. \emph{Matter} \textbf{8}, 102263 (2025). \href{https://doi.org/10.1016/j.matt.2025.102263}{DOI}
\par
\hypertarget{ref24}{}\hangindent=2.45em\hangafter=1\noindent\makebox[2.45em][l]{24.}Pérez-Sánchez, J. B. et al. El Agente Quntur: A research collaborator agent for quantum chemistry. \emph{Chem}, 103259 (2026). \href{https://doi.org/10.1016/j.chempr.2026.103259}{DOI}
\par
\hypertarget{ref25}{}\hangindent=2.45em\hangafter=1\noindent\makebox[2.45em][l]{25.}Choi, C. et al. El Agente Estructural: An Artificially Intelligent Molecular Editor. arXiv:2602.04849 (2026). \href{https://arxiv.org/abs/2602.04849}{arXiv}
\par
\hypertarget{ref26}{}\hangindent=2.45em\hangafter=1\noindent\makebox[2.45em][l]{26.}Hari Kumar, S. G. et al. El Agente Sólido: A New Age(nt) for Solid State Simulations. arXiv:2602.17886 (2026). \href{https://arxiv.org/abs/2602.17886}{arXiv}
\par
\hypertarget{ref27}{}\hangindent=2.45em\hangafter=1\noindent\makebox[2.45em][l]{27.}Ko, T. W. et al. El Agente Potente: High-Throughput Agentic Atomistic Simulations. arXiv:2609.14840 (2026). \href{https://arxiv.org/abs/2609.14840}{arXiv}
\par
\hypertarget{ref28}{}\hangindent=2.45em\hangafter=1\noindent\makebox[2.45em][l]{28.}Gustin, I. et al. El agente cuántico: automating quantum simulations. \emph{Reports on Progress in Physics} \textbf{89}, 077602 (2026). \href{https://doi.org/10.1088/1361-6633/ae8933}{DOI}
\par
\hypertarget{ref29}{}\hangindent=2.45em\hangafter=1\noindent\makebox[2.45em][l]{29.}Zhang, Z. et al. El Agente Forjador: Task-Driven Agent Generation for Quantum Simulation. arXiv:2604.14609 (2026). \href{https://arxiv.org/abs/2604.14609}{arXiv}
\par
\hypertarget{ref30}{}\hangindent=2.45em\hangafter=1\noindent\makebox[2.45em][l]{30.}Müller, M. et al. El Agente Fármaco: A Scalable AI Scientist for Drug Discovery. arXiv:2611.03832 (2026). \href{https://arxiv.org/abs/2611.03832}{arXiv}
\par
\hypertarget{ref31}{}\hangindent=2.45em\hangafter=1\noindent\makebox[2.45em][l]{31.}Pérez-Sánchez, J. B. et al. El Agente Quntur: Agentic Active-Space Selection for CASSCF. arXiv:2610.38145 (2026). \href{https://arxiv.org/abs/2610.38145}{arXiv}
\par
\hypertarget{ref32}{}\hangindent=2.45em\hangafter=1\noindent\makebox[2.45em][l]{32.}Lo, A. et al. Fast organic crystal structure prediction with unit cell flow matching. Preprint at arXiv (2026). \href{https://arxiv.org/abs/2606.03199}{Preprint}
\par
\hypertarget{ref33}{}\hangindent=2.45em\hangafter=1\noindent\makebox[2.45em][l]{33.}Wood, B. et al. UMA: A Family of Universal Models for Atoms. \emph{Advances in Neural Information Processing Systems 38}, 143528--143564 (2025). \href{https://doi.org/10.52202/085713-4310}{DOI}
\par
\hypertarget{ref34}{}\hangindent=2.45em\hangafter=1\noindent\makebox[2.45em][l]{34.}Kazmierczak, N. P., Lopez, N. E., Luedecke, K. M. \& Hadt, R. G. Determining the key vibrations for spin relaxation in ruffled Cu(II) porphyrins via resonance Raman spectroscopy. \emph{Chemical Science} \textbf{15}, 2380--2390 (2024). \href{https://doi.org/10.1039/d3sc05774g}{DOI}
\par
\hypertarget{ref35}{}\hangindent=2.45em\hangafter=1\noindent\makebox[2.45em][l]{35.}Posavec, L. \& Cinčić, D. Isothiocyanate Sulfur Atom as an Acceptor Site for Halogen-Bonded Cocrystallization of Werner Ni(II) Coordination Compounds and Perfluorinated Iodobenzenes. \emph{Crystal Growth \& Design} \textbf{24}, 7514--7523 (2024). \href{https://doi.org/10.1021/acs.cgd.4c00697}{DOI}
\par
\hypertarget{ref36}{}\hangindent=2.45em\hangafter=1\noindent\makebox[2.45em][l]{36.}Joshi, H. K. et al. Investigation of metal--dithiolate fold angle effects: Implications for molybdenum and tungsten enzymes. \emph{Proceedings of the National Academy of Sciences} \textbf{100}, 3719--3724 (2003). \href{https://doi.org/10.1073/pnas.0636832100}{DOI}
\par
\hypertarget{ref37}{}\hangindent=2.45em\hangafter=1\noindent\makebox[2.45em][l]{37.}Cranswick, M. A. et al. Photoelectron Spectroscopy and Electronic Structure Calculations of d1 Vanadocene Compounds with Chelated Dithiolate Ligands: Implications for Pyranopterin Mo/W Enzymes. \emph{Inorganic Chemistry} \textbf{46}, 10639--10646 (2007). \href{https://doi.org/10.1021/ic701338s}{DOI}
\par
\hypertarget{ref38}{}\hangindent=2.45em\hangafter=1\noindent\makebox[2.45em][l]{38.}Stein, B. W. et al. Vibrational Control of Covalency Effects Related to the Active Sites of Molybdenum Enzymes. \emph{Journal of the American Chemical Society} \textbf{140}, 14777--14788 (2018). \href{https://doi.org/10.1021/jacs.8b08254}{DOI}
\par
\hypertarget{ref39}{}\hangindent=2.45em\hangafter=1\noindent\makebox[2.45em][l]{39.}Zadrozny, J. M., Gallagher, A. T., Harris, T. D. \& Freedman, D. E. A Porous Array of Clock Qubits. \emph{Journal of the American Chemical Society} \textbf{139}, 7089--7094 (2017). \href{https://doi.org/10.1021/jacs.7b03123}{DOI}
\par
\hypertarget{ref40}{}\hangindent=2.45em\hangafter=1\noindent\makebox[2.45em][l]{40.}Yamabayashi, T. et al. Scaling Up Electronic Spin Qubits into a Three-Dimensional Metal--Organic Framework. \emph{Journal of the American Chemical Society} \textbf{140}, 12090--12101 (2018). \href{https://doi.org/10.1021/jacs.8b06733}{DOI}
\par
\hypertarget{ref41}{}\hangindent=2.45em\hangafter=1\noindent\makebox[2.45em][l]{41.}Yu, C.-J., Krzyaniak, M. D., Fataftah, M. S., Wasielewski, M. R. \& Freedman, D. E. A concentrated array of copper porphyrin candidate qubits. \emph{Chemical Science} \textbf{10}, 1702--1708 (2019). \href{https://doi.org/10.1039/c8sc04435j}{DOI}
\par
\hypertarget{ref42}{}\hangindent=2.45em\hangafter=1\noindent\makebox[2.45em][l]{42.}Oanta, A. K. et al. Electronic Spin Qubit Candidates Arrayed within Layered Two-Dimensional Polymers. \emph{Journal of the American Chemical Society} \textbf{145}, 689--696 (2023). \href{https://doi.org/10.1021/jacs.2c11784}{DOI}
\par
\hypertarget{ref43}{}\hangindent=2.45em\hangafter=1\noindent\makebox[2.45em][l]{43.}Moisanu, C. M. et al. Crystalline Arrays of Copper Porphyrin Qubits Based on Ion-Paired Frameworks. \emph{Journal of the American Chemical Society} \textbf{145}, 18447--18454 (2023). \href{https://doi.org/10.1021/jacs.3c04786}{DOI}
\par
\hypertarget{ref44}{}\hangindent=2.45em\hangafter=1\noindent\makebox[2.45em][l]{44.}Graham, M. J., Zadrozny, J. M., Fataftah, M. S. \& Freedman, D. E. Forging Solid-State Qubit Design Principles in a Molecular Furnace. \emph{Chemistry of Materials} \textbf{29}, 1885--1897 (2017). \href{https://doi.org/10.1021/acs.chemmater.6b05433}{DOI}
\par
\hypertarget{ref45}{}\hangindent=2.45em\hangafter=1\noindent\makebox[2.45em][l]{45.}Yu, C.-J. et al. Spin and Phonon Design in Modular Arrays of Molecular Qubits. \emph{Chemistry of Materials} \textbf{32}, 10200--10206 (2020). \href{https://doi.org/10.1021/acs.chemmater.0c03718}{DOI}
\par
\hypertarget{ref46}{}\hangindent=2.45em\hangafter=1\noindent\makebox[2.45em][l]{46.}Yamauchi, A. et al. Modulation of triplet quantum coherence by guest-induced structural changes in a flexible metal-organic framework. \emph{Nature Communications} \textbf{15}, 7622 (2024). \href{https://doi.org/10.1038/s41467-024-51715-w}{DOI}
\par
\hypertarget{ref47}{}\hangindent=2.45em\hangafter=1\noindent\makebox[2.45em][l]{47.}Vujević, L. et al. Improving the molecular spin qubit performance in zirconium MOF composites by mechanochemical dilution and fullerene encapsulation. \emph{Chemical Science} \textbf{14}, 9389--9399 (2023). \href{https://doi.org/10.1039/d3sc03089j}{DOI}
\par
\hypertarget{ref48}{}\hangindent=2.45em\hangafter=1\noindent\makebox[2.45em][l]{48.}Zhou, A. et al. Phononic modulation of spin-lattice relaxation in molecular qubit frameworks. \emph{Nature Communications} \textbf{15}, 10763 (2024). \href{https://doi.org/10.1038/s41467-024-54989-2}{DOI}
\par
\hypertarget{ref49}{}\hangindent=2.45em\hangafter=1\noindent\makebox[2.45em][l]{49.}Asher, M. et al. Mechanistic View on the Order-Disorder Phase Transition in Amphidynamic Crystals. \emph{The Journal of Physical Chemistry Letters} \textbf{14}, 1570--1577 (2023). \href{https://doi.org/10.1021/acs.jpclett.2c03316}{DOI}
\par
\hypertarget{ref50}{}\hangindent=2.45em\hangafter=1\noindent\makebox[2.45em][l]{50.}Nabi, R. et al. Accurate and Efficient Spin--Phonon Coupling and Spin Dynamics Calculations for Molecular Solids. \emph{Journal of the American Chemical Society} \textbf{145}, 24558--24567 (2023). \href{https://doi.org/10.1021/jacs.3c06015}{DOI}
\par
\hypertarget{ref51}{}\hangindent=2.45em\hangafter=1\noindent\makebox[2.45em][l]{51.}Moseley, D. H. et al. Spin--phonon couplings in transition metal complexes with slow magnetic relaxation. \emph{Nature Communications} \textbf{9}, 2572 (2018). \href{https://doi.org/10.1038/s41467-018-04896-0}{DOI}
\par
\hypertarget{ref52}{}\hangindent=2.45em\hangafter=1\noindent\makebox[2.45em][l]{52.}Garlatti, E. et al. The critical role of ultra-low-energy vibrations in the relaxation dynamics of molecular qubits. \emph{Nature Communications} \textbf{14}, 1653 (2023). \href{https://doi.org/10.1038/s41467-023-36852-y}{DOI}
\par
\hypertarget{ref53}{}\hangindent=2.45em\hangafter=1\noindent\makebox[2.45em][l]{53.}Kragskow, J. G. C. et al. Analysis of vibronic coupling in a 4f molecular magnet with FIRMS. \emph{Nature Communications} \textbf{13}, 825 (2022). \href{https://doi.org/10.1038/s41467-022-28352-2}{DOI}
\par
\hypertarget{ref54}{}\hangindent=2.45em\hangafter=1\noindent\makebox[2.45em][l]{54.}Kazmierczak, N. P. \& Hadt, R. G. Illuminating Ligand Field Contributions to Molecular Qubit Spin Relaxation via $T_1$ Anisotropy. \emph{Journal of the American Chemical Society} \textbf{144}, 20804--20814 (2022). \href{https://doi.org/10.1021/jacs.2c08729}{DOI}
\par
\hypertarget{ref55}{}\hangindent=2.45em\hangafter=1\noindent\makebox[2.45em][l]{55.}Kazmierczak, N. P., Luedecke, K. M., Gallmeier, E. T. \& Hadt, R. G. $T_1$ Anisotropy Elucidates Spin Relaxation Mechanisms in an S = 1 Cr(IV) Optically Addressable Molecular Qubit. \emph{The Journal of Physical Chemistry Letters} \textbf{14}, 7658--7664 (2023). \href{https://doi.org/10.1021/acs.jpclett.3c01964}{DOI}
\par
\hypertarget{ref56}{}\hangindent=2.45em\hangafter=1\noindent\makebox[2.45em][l]{56.}Lenz, S., Kern, B., Schneider, M. \& van Slageren, J. Measurement of quantum coherence in thin films of molecular quantum bits without post-processing. \emph{Chemical Communications} \textbf{55}, 7163--7166 (2019). \href{https://doi.org/10.1039/c9cc02184a}{DOI}
\par
\hypertarget{ref57}{}\hangindent=2.45em\hangafter=1\noindent\makebox[2.45em][l]{57.}Dragone, V., Sans, V., Henson, A. B., Granda, J. M. \& Cronin, L. An autonomous organic reaction search engine for chemical reactivity. \emph{Nature Communications} \textbf{8}, 15733 (2017). \href{https://doi.org/10.1038/ncomms15733}{DOI}
\par
\hypertarget{ref58}{}\hangindent=2.45em\hangafter=1\noindent\makebox[2.45em][l]{58.}Granda, J. M., Donina, L., Dragone, V., Long, D.-L. \& Cronin, L. Controlling an organic synthesis robot with machine learning to search for new reactivity. \emph{Nature} \textbf{559}, 377--381 (2018). \href{https://doi.org/10.1038/s41586-018-0307-8}{DOI}
\par
\hypertarget{ref59}{}\hangindent=2.45em\hangafter=1\noindent\makebox[2.45em][l]{59.}Caramelli, D. et al. Discovering New Chemistry with an Autonomous Robotic Platform Driven by a Reactivity-Seeking Neural Network. \emph{ACS Central Science} \textbf{7}, 1821--1830 (2021). \href{https://doi.org/10.1021/acscentsci.1c00435}{DOI}
\par
\hypertarget{ref60}{}\hangindent=2.45em\hangafter=1\noindent\makebox[2.45em][l]{60.}Shields, B. J. et al. Bayesian reaction optimization as a tool for chemical synthesis. \emph{Nature} \textbf{590}, 89--96 (2021). \href{https://doi.org/10.1038/s41586-021-03213-y}{DOI}
\par
\hypertarget{ref61}{}\hangindent=2.45em\hangafter=1\noindent\makebox[2.45em][l]{61.}Angello, N. H. et al. Closed-loop optimization of general reaction conditions for heteroaryl Suzuki-Miyaura coupling. \emph{Science} \textbf{378}, 399--405 (2022). \href{https://doi.org/10.1126/science.adc8743}{DOI}
\par
\hypertarget{ref62}{}\hangindent=2.45em\hangafter=1\noindent\makebox[2.45em][l]{62.}MacLeod, B. P. et al. Self-driving laboratory for accelerated discovery of thin-film materials. \emph{Science Advances} \textbf{6}, eaaz8867 (2020). \href{https://doi.org/10.1126/sciadv.aaz8867}{DOI}
\par
\hypertarget{ref63}{}\hangindent=2.45em\hangafter=1\noindent\makebox[2.45em][l]{63.}Strieth-Kalthoff, F. et al. Delocalized, asynchronous, closed-loop discovery of organic laser emitters. \emph{Science} \textbf{384}, eadk9227 (2024). \href{https://doi.org/10.1126/science.adk9227}{DOI}
\par
\hypertarget{ref64}{}\hangindent=2.45em\hangafter=1\noindent\makebox[2.45em][l]{64.}Koscher, B. A. et al. Autonomous, multiproperty-driven molecular discovery: From predictions to measurements and back. \emph{Science} \textbf{382}, eadi1407 (2023). \href{https://doi.org/10.1126/science.adi1407}{DOI}
\par
\hypertarget{ref65}{}\hangindent=2.45em\hangafter=1\noindent\makebox[2.45em][l]{65.}Szymanski, N. J. et al. An autonomous laboratory for the accelerated synthesis of inorganic materials. \emph{Nature} \textbf{624}, 86--91 (2023). \href{https://doi.org/10.1038/s41586-023-06734-w}{DOI}
\par
\hypertarget{ref66}{}\hangindent=2.45em\hangafter=1\noindent\makebox[2.45em][l]{66.}Szymanski, N. J., Nevatia, P., Bartel, C. J., Zeng, Y. \& Ceder, G. Autonomous and dynamic precursor selection for solid-state materials synthesis. \emph{Nature Communications} \textbf{14}, 6956 (2023). \href{https://doi.org/10.1038/s41467-023-42329-9}{DOI}
\par
\hypertarget{ref67}{}\hangindent=2.45em\hangafter=1\noindent\makebox[2.45em][l]{67.}Szymanski, N. J. et al. Adaptively driven X-ray diffraction guided by machine learning for autonomous phase identification. \emph{npj Computational Materials} \textbf{9}, 31 (2023). \href{https://doi.org/10.1038/s41524-023-00984-y}{DOI}
\par
\hypertarget{ref68}{}\hangindent=2.45em\hangafter=1\noindent\makebox[2.45em][l]{68.}Gustin, I., Kim, C. W., McCamant, D. W. \& Franco, I. Mapping electronic decoherence pathways in molecules. \emph{Proceedings of the National Academy of Sciences} \textbf{120}, e2309987120 (2023). \href{https://doi.org/10.1073/pnas.2309987120}{DOI}
\par
\hypertarget{ref69}{}\hangindent=2.45em\hangafter=1\noindent\makebox[2.45em][l]{69.}Burger, B. et al. A mobile robotic chemist. \emph{Nature} \textbf{583}, 237--241 (2020). \href{https://doi.org/10.1038/s41586-020-2442-2}{DOI}
\par
\hypertarget{ref70}{}\hangindent=2.45em\hangafter=1\noindent\makebox[2.45em][l]{70.}Coley, C. W. et al. A robotic platform for flow synthesis of organic compounds informed by AI planning. \emph{Science} \textbf{365}, eaax1566 (2019). \href{https://doi.org/10.1126/science.aax1566}{DOI}
\par
\hypertarget{ref71}{}\hangindent=2.45em\hangafter=1\noindent\makebox[2.45em][l]{71.}Darvish, K. et al. ORGANA: A robotic assistant for automated chemistry experimentation and characterization. \emph{Matter} \textbf{8}, 101897 (2025). \href{https://doi.org/10.1016/j.matt.2024.10.015}{DOI}
\par
\hypertarget{ref72}{}\hangindent=2.45em\hangafter=1\noindent\makebox[2.45em][l]{72.}Ruan, Y. et al. An automatic end-to-end chemical synthesis development platform powered by large language models. \emph{Nature Communications} \textbf{15}, 10160 (2024). \href{https://doi.org/10.1038/s41467-024-54457-x}{DOI}
\par
\hypertarget{ref73}{}\hangindent=2.45em\hangafter=1\noindent\makebox[2.45em][l]{73.}Dai, T. et al. Autonomous mobile robots for exploratory synthetic chemistry. \emph{Nature} \textbf{635}, 890--897 (2024). \href{https://doi.org/10.1038/s41586-024-08173-7}{DOI}
\par
\hypertarget{ref74}{}\hangindent=2.45em\hangafter=1\noindent\makebox[2.45em][l]{74.}Song, T. et al. A Multiagent-Driven Robotic AI Chemist Enabling Autonomous Chemical Research On Demand. \emph{Journal of the American Chemical Society} \textbf{147}, 12534--12545 (2025). \href{https://doi.org/10.1021/jacs.4c17738}{DOI}
\par
\hypertarget{ref75}{}\hangindent=2.45em\hangafter=1\noindent\makebox[2.45em][l]{75.}Aspuru-Guzik, A. \& Bernales, V. The rise of agents: Computational chemistry is ready for (R)evolution. \emph{Polyhedron} \textbf{281}, 117707 (2025). \href{https://doi.org/10.1016/j.poly.2025.117707}{DOI}
\par
\hypertarget{ref76}{}\hangindent=2.45em\hangafter=1\noindent\makebox[2.45em][l]{76.}Feng, B. et al. Agentic AI systems for scientific discovery. \emph{Nature Reviews Methods Primers} \textbf{6}, 81 (2026). \href{https://doi.org/10.1038/s43586-026-00534-y}{DOI}
\par
\hypertarget{ref77}{}\hangindent=2.45em\hangafter=1\noindent\makebox[2.45em][l]{77.}M. Bran, A. et al. Augmenting large language models with chemistry tools. \emph{Nature Machine Intelligence} \textbf{6}, 525--535 (2024). \href{https://doi.org/10.1038/s42256-024-00832-8}{DOI}
\par
\hypertarget{ref78}{}\hangindent=2.45em\hangafter=1\noindent\makebox[2.45em][l]{78.}Ding, K. et al. SciToolAgent: a knowledge-graph-driven scientific agent for multitool integration. \emph{Nature Computational Science} \textbf{5}, 962--972 (2025). \href{https://doi.org/10.1038/s43588-025-00849-y}{DOI}
\par
\hypertarget{ref79}{}\hangindent=2.45em\hangafter=1\noindent\makebox[2.45em][l]{79.}McNaughton, A. D. et al. CACTUS: Chemistry Agent Connecting Tool Usage to Science. \emph{ACS Omega} \textbf{9}, 46563--46573 (2024). \href{https://doi.org/10.1021/acsomega.4c08408}{DOI}
\par
\hypertarget{ref80}{}\hangindent=2.45em\hangafter=1\noindent\makebox[2.45em][l]{80.}Ansari, M. \& Moosavi, S. M. Agent-based learning of materials datasets from the scientific literature. \emph{Digital Discovery} \textbf{3}, 2607--2617 (2024). \href{https://doi.org/10.1039/d4dd00252k}{DOI}
\par
\hypertarget{ref81}{}\hangindent=2.45em\hangafter=1\noindent\makebox[2.45em][l]{81.}Ma, K. AI agents in chemical research: GVIM -- an intelligent research assistant system. \emph{Digital Discovery} \textbf{4}, 355--375 (2025). \href{https://doi.org/10.1039/d4dd00398e}{DOI}
\par
\hypertarget{ref82}{}\hangindent=2.45em\hangafter=1\noindent\makebox[2.45em][l]{82.}Ghafarollahi, A. \& Buehler, M. J. Automating alloy design and discovery with physics-aware multimodal multiagent AI. \emph{Proceedings of the National Academy of Sciences} \textbf{122}, e2414074122 (2025). \href{https://doi.org/10.1073/pnas.2414074122}{DOI}
\par
\hypertarget{ref83}{}\hangindent=2.45em\hangafter=1\noindent\makebox[2.45em][l]{83.}Ghafarollahi, A. \& Buehler, M. J. Autonomous in-silico inorganic materials discovery via multi-agent physics-aware scientific reasoning. \emph{npj Computational Materials} \textbf{12}, 357 (2026). \href{https://doi.org/10.1038/s41524-026-02205-8}{DOI}
\par
\hypertarget{ref84}{}\hangindent=2.45em\hangafter=1\noindent\makebox[2.45em][l]{84.}Kang, Y. \& Kim, J. ChatMOF: an artificial intelligence system for predicting and generating metal-organic frameworks using large language models. \emph{Nature Communications} \textbf{15}, 4705 (2024). \href{https://doi.org/10.1038/s41467-024-48998-4}{DOI}
\par
\hypertarget{ref85}{}\hangindent=2.45em\hangafter=1\noindent\makebox[2.45em][l]{85.}Campbell, Q., Cox, S., Medina, J., Watterson, B. \& White, A. D. MDCrow: automating molecular dynamics workflows with large language models. \emph{Machine Learning: Science and Technology} \textbf{7}, 025037 (2026). \href{https://doi.org/10.1088/2632-2153/ae4b07}{DOI}
\par
\hypertarget{ref86}{}\hangindent=2.45em\hangafter=1\noindent\makebox[2.45em][l]{86.}Liu, J. et al. VASPilot: MCP-facilitated multi-agent intelligence for autonomous VASP simulations. \emph{Chinese Physics B} \textbf{34}, 117106 (2025). \href{https://doi.org/10.1088/1674-1056/ae0681}{DOI}
\par
\hypertarget{ref87}{}\hangindent=2.45em\hangafter=1\noindent\makebox[2.45em][l]{87.}Pham, T. D., Tanikanti, A. \& Keçeli, M. ChemGraph as an agentic framework for computational chemistry workflows. \emph{Communications Chemistry} \textbf{9}, 33 (2026). \href{https://doi.org/10.1038/s42004-025-01776-9}{DOI}
\par
\hypertarget{ref88}{}\hangindent=2.45em\hangafter=1\noindent\makebox[2.45em][l]{88.}Vriza, A., Kornu, U., Koneru, A., Chan, H. \& Sankaranarayanan, S. K. R. S. Multi-agentic AI framework for end-to-end atomistic simulations. \emph{Digital Discovery} \textbf{5}, 440--452 (2026). \href{https://doi.org/10.1039/d5dd00435g}{DOI}
\par
\hypertarget{ref89}{}\hangindent=2.45em\hangafter=1\noindent\makebox[2.45em][l]{89.}Soleymanibrojeni, M. et al. GENIUS: an agentic AI framework for autonomous design and execution of simulation protocols. \emph{Communications Materials} \textbf{7}, 115 (2026). \href{https://doi.org/10.1038/s43246-026-01167-0}{DOI}
\par
\hypertarget{ref90}{}\hangindent=2.45em\hangafter=1\noindent\makebox[2.45em][l]{90.}Yang, F. \& Evans, J. D. QUASAR: A Universal Autonomous System for Atomistic Simulation and a Benchmark of Its Capabilities. \emph{Journal of Chemical Information and Modeling} \textbf{66}, 5911--5918 (2026). \href{https://doi.org/10.1021/acs.jcim.6c00295}{DOI}
\par
\hypertarget{ref91}{}\hangindent=2.45em\hangafter=1\noindent\makebox[2.45em][l]{91.}Curtarolo, S. et al. AFLOW: An automatic framework for high-throughput materials discovery. \emph{Computational Materials Science} \textbf{58}, 218--226 (2012). \href{https://doi.org/10.1016/j.commatsci.2012.02.005}{DOI}
\par
\hypertarget{ref92}{}\hangindent=2.45em\hangafter=1\noindent\makebox[2.45em][l]{92.}Pizzi, G., Cepellotti, A., Sabatini, R., Marzari, N. \& Kozinsky, B. AiiDA: automated interactive infrastructure and database for computational science. \emph{Computational Materials Science} \textbf{111}, 218--230 (2016). \href{https://doi.org/10.1016/j.commatsci.2015.09.013}{DOI}
\par
\hypertarget{ref93}{}\hangindent=2.45em\hangafter=1\noindent\makebox[2.45em][l]{93.}Graham, M. J., Yu, C.-J., Krzyaniak, M. D., Wasielewski, M. R. \& Freedman, D. E. Synthetic Approach To Determine the Effect of Nuclear Spin Distance on Electronic Spin Decoherence. \emph{Journal of the American Chemical Society} \textbf{139}, 3196--3201 (2017). \href{https://doi.org/10.1021/jacs.6b13030}{DOI}
\par
\hypertarget{ref94}{}\hangindent=2.45em\hangafter=1\noindent\makebox[2.45em][l]{94.}Kühne, T. D. et al. CP2K: An electronic structure and molecular dynamics software package - Quickstep: Efficient and accurate electronic structure calculations. \emph{The Journal of Chemical Physics} \textbf{152}, 194103 (2020). \href{https://doi.org/10.1063/5.0007045}{DOI}
\par
\hypertarget{ref95}{}\hangindent=2.45em\hangafter=1\noindent\makebox[2.45em][l]{95.}Perdew, J. P., Burke, K. \& Ernzerhof, M. Generalized Gradient Approximation Made Simple. \emph{Physical Review Letters} \textbf{77}, 3865--3868 (1996). \href{https://doi.org/10.1103/physrevlett.77.3865}{DOI}
\par
\hypertarget{ref96}{}\hangindent=2.45em\hangafter=1\noindent\makebox[2.45em][l]{96.}Grimme, S., Antony, J., Ehrlich, S. \& Krieg, H. A consistent and accurate ab initio parametrization of density functional dispersion correction (DFT-D) for the 94 elements H-Pu. \emph{The Journal of Chemical Physics} \textbf{132}, 154104 (2010). \href{https://doi.org/10.1063/1.3382344}{DOI}
\par
\hypertarget{ref97}{}\hangindent=2.45em\hangafter=1\noindent\makebox[2.45em][l]{97.}VandeVondele, J. \& Hutter, J. Gaussian basis sets for accurate calculations on molecular systems in gas and condensed phases. \emph{The Journal of Chemical Physics} \textbf{127}, 114105 (2007). \href{https://doi.org/10.1063/1.2770708}{DOI}
\par
\hypertarget{ref98}{}\hangindent=2.45em\hangafter=1\noindent\makebox[2.45em][l]{98.}Goedecker, S., Teter, M. \& Hutter, J. Separable dual-space Gaussian pseudopotentials. \emph{Physical Review B} \textbf{54}, 1703--1710 (1996). \href{https://doi.org/10.1103/physrevb.54.1703}{DOI}
\par
\hypertarget{ref99}{}\hangindent=2.45em\hangafter=1\noindent\makebox[2.45em][l]{99.}Hartwigsen, C., Goedecker, S. \& Hutter, J. Relativistic separable dual-space Gaussian pseudopotentials from H to Rn. \emph{Physical Review B} \textbf{58}, 3641--3662 (1998). \href{https://doi.org/10.1103/physrevb.58.3641}{DOI}
\par
\hypertarget{ref100}{}\hangindent=2.45em\hangafter=1\noindent\makebox[2.45em][l]{100.}Togo, A. \& Tanaka, I. First principles phonon calculations in materials science. \emph{Scripta Materialia} \textbf{108}, 1--5 (2015). \href{https://doi.org/10.1016/j.scriptamat.2015.07.021}{DOI}
\par
\hypertarget{ref101}{}\hangindent=2.45em\hangafter=1\noindent\makebox[2.45em][l]{101.}Lunghi, A. Toward exact predictions of spin-phonon relaxation times: An ab initio implementation of open quantum systems theory. \emph{Science Advances} \textbf{8}, eabn7880 (2022). \href{https://doi.org/10.1126/sciadv.abn7880}{DOI}
\par
\hypertarget{ref102}{}\hangindent=2.45em\hangafter=1\noindent\makebox[2.45em][l]{102.}Reta, D., Kragskow, J. G. C. \& Chilton, N. F. Ab Initio Prediction of High-Temperature Magnetic Relaxation Rates in Single-Molecule Magnets. \emph{Journal of the American Chemical Society} \textbf{143}, 5943--5950 (2021). \href{https://doi.org/10.1021/jacs.1c01410}{DOI}
\par
\hypertarget{ref103}{}\hangindent=2.45em\hangafter=1\noindent\makebox[2.45em][l]{103.}Lunghi, A. \& Sanvito, S. Multiple spin--phonon relaxation pathways in a Kramer single-ion magnet. \emph{The Journal of Chemical Physics} \textbf{153}, 174113 (2020). \href{https://doi.org/10.1063/5.0017118}{DOI}
\par
\hypertarget{ref104}{}\hangindent=2.45em\hangafter=1\noindent\makebox[2.45em][l]{104.}Briganti, M. et al. A Complete Ab Initio View of Orbach and Raman Spin--Lattice Relaxation in a Dysprosium Coordination Compound. \emph{Journal of the American Chemical Society} \textbf{143}, 13633--13645 (2021). \href{https://doi.org/10.1021/jacs.1c05068}{DOI}
\par
\hypertarget{ref105}{}\hangindent=2.45em\hangafter=1\noindent\makebox[2.45em][l]{105.}Haldar, S., Mariano, L. A., Lunghi, A. \& Gagliardi, L. Role of Electron Correlation beyond the Active Space in Achieving Quantitative Predictions of Spin-Phonon Relaxation. \emph{Journal of Chemical Theory and Computation} \textbf{21}, 2829--2838 (2025). \href{https://doi.org/10.1021/acs.jctc.4c01696}{DOI}
\par
\hypertarget{ref106}{}\hangindent=2.45em\hangafter=1\noindent\makebox[2.45em][l]{106.}Verma, S. et al. Multireference Density Matrix Embedding for Spin-Phonon Relaxation. \emph{Journal of Chemical Theory and Computation} \textbf{22}, 10732--10745 (2026). \href{https://doi.org/10.1021/acs.jctc.6c01503}{DOI}
\par
\hypertarget{ref107}{}\hangindent=2.45em\hangafter=1\noindent\makebox[2.45em][l]{107.}Briganti, V. \& Lunghi, A. A machine-learning framework for accelerating spin-lattice relaxation simulations. \emph{npj Computational Materials} \textbf{11}, 62 (2025). \href{https://doi.org/10.1038/s41524-025-01547-z}{DOI}
\par
\hypertarget{ref108}{}\hangindent=2.45em\hangafter=1\noindent\makebox[2.45em][l]{108.}Neese, F. Software Update: The ORCA Program System---Version 6.0. \emph{WIREs Computational Molecular Science} \textbf{15}, e70019 (2025). \href{https://doi.org/10.1002/wcms.70019}{DOI}
\par
\hypertarget{ref109}{}\hangindent=2.45em\hangafter=1\noindent\makebox[2.45em][l]{109.}Roos, B. O., Taylor, P. R. \& Sigbahn, P. E. M. A complete active space SCF method (CASSCF) using a density matrix formulated super-CI approach. \emph{Chemical Physics} \textbf{48}, 157--173 (1980). \href{https://doi.org/10.1016/0301-0104(80)80045-0}{DOI}
\par
\hypertarget{ref110}{}\hangindent=2.45em\hangafter=1\noindent\makebox[2.45em][l]{110.}Angeli, C., Cimiraglia, R. \& Malrieu, J.-P. n-electron valence state perturbation theory: A spinless formulation and an efficient implementation of the strongly contracted and of the partially contracted variants. \emph{The Journal of Chemical Physics} \textbf{117}, 9138--9153 (2002). \href{https://doi.org/10.1063/1.1515317}{DOI}
\par
\hypertarget{ref111}{}\hangindent=2.45em\hangafter=1\noindent\makebox[2.45em][l]{111.}Neese, F. Efficient and accurate approximations to the molecular spin-orbit coupling operator and their use in molecular g-tensor calculations. \emph{The Journal of Chemical Physics} \textbf{122}, 034107 (2005). \href{https://doi.org/10.1063/1.1829047}{DOI}
\par
\hypertarget{ref112}{}\hangindent=2.45em\hangafter=1\noindent\makebox[2.45em][l]{112.}Weigend, F. \& Ahlrichs, R. Balanced basis sets of split valence, triple zeta valence and quadruple zeta valence quality for H to Rn: Design and assessment of accuracy. \emph{Physical Chemistry Chemical Physics} \textbf{7}, 3297--3305 (2005). \href{https://doi.org/10.1039/b508541a}{DOI}
\par
\hypertarget{ref113}{}\hangindent=2.45em\hangafter=1\noindent\makebox[2.45em][l]{113.}Atanasov, M., Ganyushin, D., Sivalingam, K. \& Neese, F. A Modern First-Principles View on Ligand Field Theory Through the Eyes of Correlated Multireference Wavefunctions. In \emph{Molecular Electronic Structures of Transition Metal Complexes II}, Structure and Bonding \textbf{143}, 149--220 (Springer, 2012). \href{https://doi.org/10.1007/430_2011_57}{DOI}
\par
\hypertarget{ref114}{}\hangindent=2.45em\hangafter=1\noindent\makebox[2.45em][l]{114.}Lohaus, S. H. et al. Quantifying the effects of peripheral substituents on the spin-lattice relaxation of a vanadyl molecular quantum bit. \emph{Journal of Porphyrins and Phthalocyanines} \textbf{28}, 383--389 (2024). \href{https://doi.org/10.1142/S1088424624500329}{DOI}
\par
\hypertarget{ref115}{}\hangindent=2.45em\hangafter=1\noindent\makebox[2.45em][l]{115.}Van Vleck, J. H. Paramagnetic Relaxation Times for Titanium and Chrome Alum. \emph{Physical Review} \textbf{57}, 426--447 (1940). \href{https://doi.org/10.1103/physrev.57.426}{DOI}
\par
\hypertarget{ref116}{}\hangindent=2.45em\hangafter=1\noindent\makebox[2.45em][l]{116.}Mattuck, R. D. \& Strandberg, M. W. P. Spin-Phonon Interaction in Paramagnetic Crystals. \emph{Physical Review} \textbf{119}, 1204--1217 (1960). \href{https://doi.org/10.1103/physrev.119.1204}{DOI}
\par
\hypertarget{ref117}{}\hangindent=2.45em\hangafter=1\noindent\makebox[2.45em][l]{117.}Goldman, M. Formal Theory of Spin--Lattice Relaxation. \emph{Journal of Magnetic Resonance} \textbf{149}, 160--187 (2001). \href{https://doi.org/10.1006/jmre.2000.2239}{DOI}
\par
\hypertarget{ref118}{}\hangindent=2.45em\hangafter=1\noindent\makebox[2.45em][l]{118.}Orbach, R. Spin-lattice relaxation in rare-earth salts. \emph{Proceedings of the Royal Society of London. Series A. Mathematical and Physical Sciences} \textbf{264}, 458--484 (1961). \href{https://doi.org/10.1098/rspa.1961.0211}{DOI}
\par
\hypertarget{ref119}{}\hangindent=2.45em\hangafter=1\noindent\makebox[2.45em][l]{119.}Lunghi, A. \& Sanvito, S. The Limit of Spin Lifetime in Solid-State Electronic Spins. \emph{The Journal of Physical Chemistry Letters} \textbf{11}, 6273--6278 (2020). \href{https://doi.org/10.1021/acs.jpclett.0c01681}{DOI}
\par
\hypertarget{ref120}{}\hangindent=2.45em\hangafter=1\noindent\makebox[2.45em][l]{120.}Gu, L. \& Wu, R. Origin of the anomalously low Raman exponents in single molecule magnets. \emph{Physical Review B} \textbf{103}, 014401 (2021). \href{https://doi.org/10.1103/physrevb.103.014401}{DOI}
\par
\hypertarget{ref121}{}\hangindent=2.45em\hangafter=1\noindent\makebox[2.45em][l]{121.}Mariano, L. A. et al. The role of electronic excited states in the spin-lattice relaxation of spin-1/2 molecules. \emph{Science Advances} \textbf{11}, eadr0168 (2025). \href{https://doi.org/10.1126/sciadv.adr0168}{DOI}
\par
\hypertarget{ref122}{}\hangindent=2.45em\hangafter=1\noindent\makebox[2.45em][l]{122.}Albino, A. et al. First-Principles Investigation of Spin--Phonon Coupling in Vanadium-Based Molecular Spin Quantum Bits. \emph{Inorganic Chemistry} \textbf{58}, 10260--10268 (2019). \href{https://doi.org/10.1021/acs.inorgchem.9b01407}{DOI}
\par
\hypertarget{ref123}{}\hangindent=2.45em\hangafter=1\noindent\makebox[2.45em][l]{123.}Mirzoyan, R., Kazmierczak, N. P. \& Hadt, R. G. Deconvolving Contributions to Decoherence in Molecular Electron Spin Qubits: A Dynamic Ligand Field Approach. \emph{Chemistry -- A European Journal} \textbf{27}, 9482--9494 (2021). \href{https://doi.org/10.1002/chem.202100845}{DOI}
\par
\hypertarget{ref124}{}\hangindent=2.45em\hangafter=1\noindent\makebox[2.45em][l]{124.}Shushkov, P. A novel non-adiabatic spin relaxation mechanism in molecular qubits. \emph{The Journal of Chemical Physics} \textbf{160}, 164105 (2024). \href{https://doi.org/10.1063/5.0198519}{DOI}
\par
\hypertarget{ref125}{}\hangindent=2.45em\hangafter=1\noindent\makebox[2.45em][l]{125.}Lebedev, V. I. Quadratures on a sphere. \emph{USSR Computational Mathematics and Mathematical Physics} \textbf{16}, 10--24 (1976). \href{https://doi.org/10.1016/0041-5553(76)90100-2}{DOI}
\par
\hypertarget{ref126}{}\hangindent=2.45em\hangafter=1\noindent\makebox[2.45em][l]{126.}Gustin, I., Kim, C. W. \& Franco, I. General framework for quantifying dissipation pathways in open quantum systems. III. Off-diagonal subsystem-bath couplings. \emph{The Journal of Chemical Physics} \textbf{164}, 034105 (2026). \href{https://doi.org/10.1063/5.0305296}{DOI}
\par
}

\clearpage
\setcounter{equation}{0}
\setcounter{figure}{0}
\setcounter{table}{0}
\renewcommand{\theequation}{S\arabic{equation}}
\renewcommand{\thefigure}{S\arabic{figure}}
\renewcommand{\thetable}{S\arabic{table}}
\renewcommand{\theHequation}{supp.\arabic{equation}}
\renewcommand{\theHfigure}{supp.\arabic{figure}}
\renewcommand{\theHtable}{supp.\arabic{table}}
\pdfbookmark[0]{Supplementary information}{supplementary-information}

\color{ink}

\markright{\footnotesize\sffamily Supplementary information}
\thispagestyle{plain}
{\raggedright\sffamily\bfseries\fontsize{23}{26}\selectfont Supplementary information\par}\vspace{5pt}
El Agente en acci\'on: Closed-loop prediction of longer molecular qubit lifetimes
\par
{\raggedright\sffamily\bfseries\normalsize \mbox{Ignacio Gustin\textsuperscript{1,2}}, \mbox{Kahn Shah\textsuperscript{3}}, \mbox{Austin Cheng\textsuperscript{1,2,3,4}}, \mbox{Juan B. Pérez-Sánchez\textsuperscript{2}}, \mbox{Adem Chlyah\textsuperscript{5}},  \mbox{Alán Aspuru-Guzik\textsuperscript{1,2,3,4,6,7,8,9,10,*}},\mbox{Varinia Bernales\textsuperscript{2,3,9,*}}\par}
{\raggedright\footnotesize\setlength{\parskip}{1pt}
\hangindent=1.3em\hangafter=1\noindent\makebox[1.3em][l]{\textsuperscript{1}}NVIDIA, 431 King St. W \#6th, Toronto, ON M5V 1K4, Canada\par
\hangindent=1.3em\hangafter=1\noindent\makebox[1.3em][l]{\textsuperscript{2}}Department of Chemistry, University of Toronto, 80 St. George St., Toronto, ON M5S 3H6, Canada\par
\hangindent=1.3em\hangafter=1\noindent\makebox[1.3em][l]{\textsuperscript{3}}Department of Computer Science, University of Toronto, 40 St. George St., Toronto, ON M5S 2E4, Canada\par
\hangindent=1.3em\hangafter=1\noindent\makebox[1.3em][l]{\textsuperscript{4}}Vector Institute for Artificial Intelligence, W1140-108 College St., Schwartz Reisman Innovation Campus, Toronto, ON M5G 0C6, Canada\par
\hangindent=1.3em\hangafter=1\noindent\makebox[1.3em][l]{\textsuperscript{5}}Department of Mathematical and Computational Sciences, University of Toronto Mississauga, Mississauga, ON, Canada\newline
3359 Mississauga Rd, Mississauga, ON L5L 1C6\par
\hangindent=1.3em\hangafter=1\noindent\makebox[1.3em][l]{\textsuperscript{6}}Department of Materials Science \& Engineering, University of Toronto, 184 College St., Toronto, ON M5S 3E4, Canada\par
\hangindent=1.3em\hangafter=1\noindent\makebox[1.3em][l]{\textsuperscript{7}}Department of Chemical Engineering \& Applied Chemistry, University of Toronto, 200 College St., Toronto, ON M5S 3E5, Canada\par
\hangindent=1.3em\hangafter=1\noindent\makebox[1.3em][l]{\textsuperscript{8}}Acceleration Consortium, 700 University Ave., Toronto, ON M7A 2S4, Canada\par
\hangindent=1.3em\hangafter=1\noindent\makebox[1.3em][l]{\textsuperscript{9}}Canadian Institute for Advanced Research (CIFAR), 661 University Ave., Toronto, ON M5G 1M1, Canada\par
\hangindent=1.3em\hangafter=1\noindent\makebox[1.3em][l]{\textsuperscript{10}}Institute of Medical Science, 1 King’s College Circle, Medical Sciences Building, Room 2374, Toronto, ON M5S 1A8, Canada\par
\hangindent=1.3em\hangafter=1\noindent\makebox[1.3em][l]{\textsuperscript{*}}Corresponding authors: alan@aspuru.com, varinia@bernales.org\par
}
\pdfbookmark[1]{Experimental controls}{secExperimentalcontrols}
\section*{Experimental controls}
Before the search we compared calculated lifetimes with pulsed electron paramagnetic resonance measurements for three molecular spins. The vanadyl phthalocyanine benchmark used a titanium host. The other two comparisons were blind to the measured lifetimes. For copper octaethylporphyrin we used the measured $g$ tensor to set the spin--orbit scale. The source citations in Table S1 use the numbering of the main article.
\par
\medskip{\sffamily\small \textbf{Supplementary Table S1. Experimental benchmarks for the excited-state relaxation model.}\par}\nopagebreak
{\centering\small\sffamily\setlength{\tabcolsep}{3.2pt}\renewcommand{\arraystretch}{1.16}\begin{tabular*}{\linewidth}{@{}l@{\extracolsep{\fill}}ccc@{}}\toprule
\textbf{Spin centre and host} & \textbf{Measurement (K)} & \textbf{Calculated $T_1$} & \textbf{Measured $T_1$} \\ \midrule
VOPc in TiOPc \href{https://doi.org/10.1021/jacs.5b13408}{[1]} & 40 & 0.61 ms & 0.22 ms \\
VOPc in TiOPc \href{https://doi.org/10.1021/jacs.5b13408}{[1]} & 100 & 14.1 \(\mu\)s & 20 \(\mu\)s \\
VOPc in TiOPc \href{https://doi.org/10.1021/jacs.5b13408}{[1]} & 300 & 0.84 \(\mu\)s & 1.1 \(\mu\)s \\
CuOEP in ZnOEP \href{https://doi.org/10.1039/d3sc05774g}{[27]} & 30--300 & See series comparison\textsuperscript{1} & See series comparison\textsuperscript{1} \\
Vanadyl dmit parent \href{https://doi.org/10.1021/jacs.6b05574}{[3]} & 293 & 1.76 \(\mu\)s & 3.37 \(\mu\)s \\
\bottomrule\end{tabular*}\par}
\textsuperscript{1} The campaign records deviations of 1.1--1.8-fold over 18 CuOEP comparisons. The experimental study reports parallel and perpendicular field positions. The archived comparison does not identify the trace used or provide the 18 calculated values. For the vanadyl salt, the measured value is for a crystal containing 5 ± 1\% vanadyl in a diamagnetic molybdenyl host at 293 K, with the field close to the vanadyl axis. The parent reference used to rank designs was instead a powder calculation at 300 K and 0.345 T (Table S2). The separate host control contains one vanadyl and one molybdenyl anion per calculation cell.
\par
At room temperature the ground-state spin-Hamiltonian contribution accounted for less than 0.2\% of the VOPc rate. The calculated vanadyl d\(\pi\) energy agreed with the range inferred from the measured transverse g value. For the parent, extending the vibrational cutoff from 500 to 1,100 cm\(^{-1}\) changed the lifetime by a factor of 0.997, and the finite-displacement step test changed it by 0.76\%. Restarting from independent orbitals gave lifetimes within 1\% of each other.
\par
The blind parent benchmark and the candidate comparisons answer different questions. Table S1 compares the absolute lifetime with experiment. Tables S3 and S4 compare each candidate with the calculated parent at the same phonon-pair broadening, temperature, field and sampling rule.
\par
\begin{figure}[!htbp]
\pdfbookmark[1]{Supplementary figures}{suppfigs}
\section*{Supplementary figures}
\centering\includegraphics[width=\linewidth]{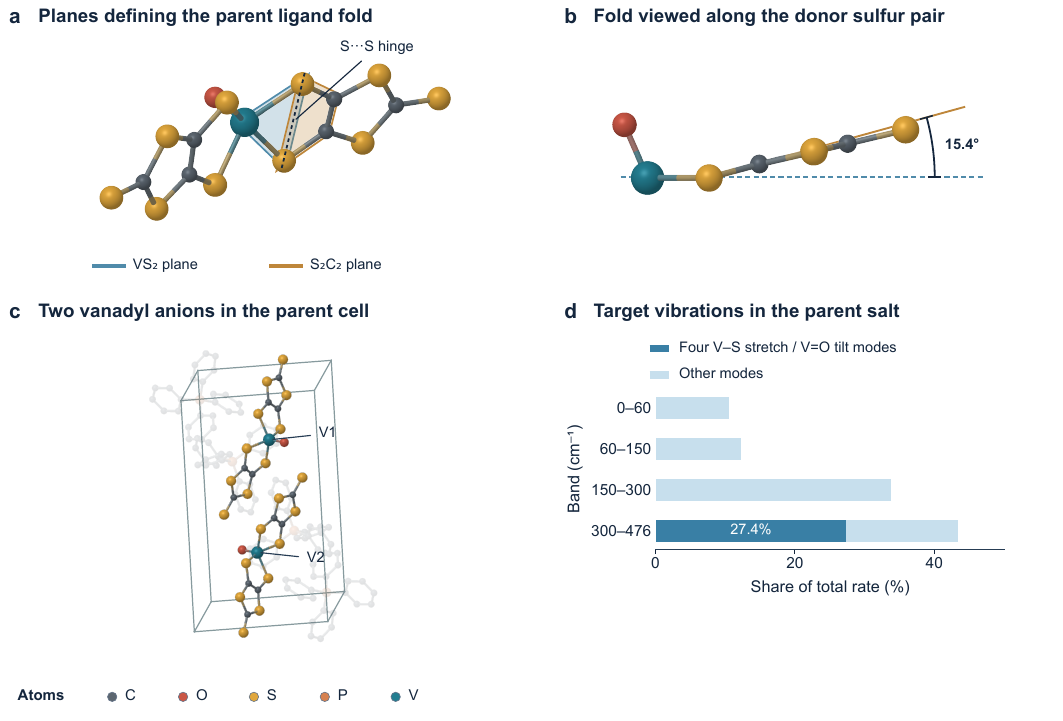}\par
{\raggedright\rmfamily\small \textbf{Supplementary Fig. S1. The ligand fold and the target vibration in the parent salt.} a,b, The blue VS$_2$ and ochre S$_2$C$_2$ planes meet at the sulfur hinge. The larger parent fold is 15.4\(^{\circ}\). c, V1 and V2 identify the spin-bearing anions. d, Zone-centre rate by frequency band at 300 K, 0.345 T and \(\sigma\) = 10 cm\(^{-1}\). Four V–S stretch and V=O tilt modes (dark blue) account for 27.4\% of the total. Atom colors: C, dark grey; O, red; S, gold; P, orange; V, teal.\par}
\end{figure}
\begin{figure}[!htbp]
\centering\includegraphics[width=\linewidth]{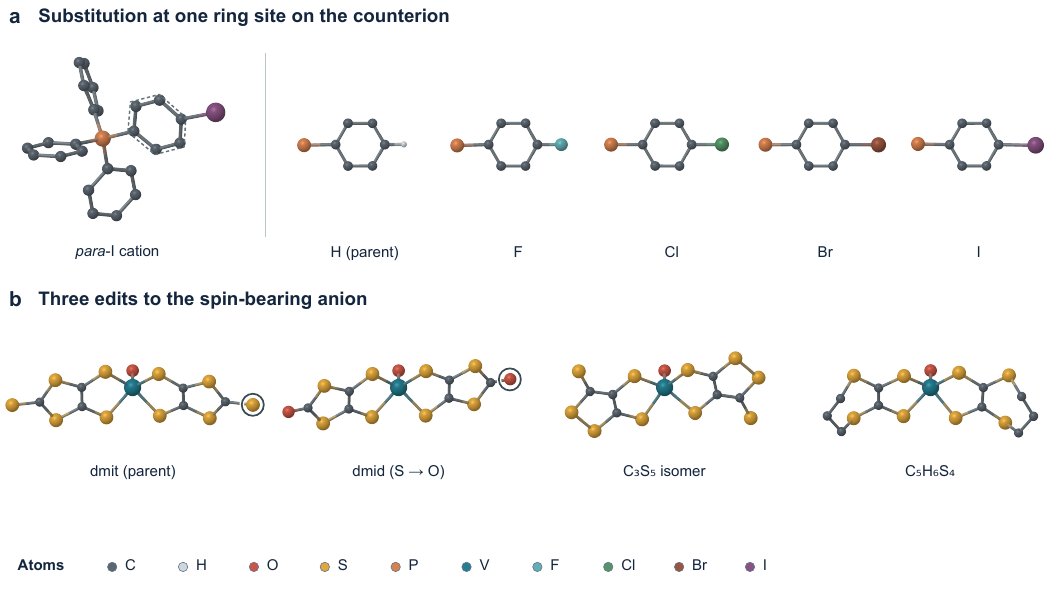}\par
{\raggedright\rmfamily\small \textbf{Supplementary Fig. S2. The edited aryl group and the four vanadyl ligand structures.} a, The outline identifies the substituted ring on the iodine counterion. Enlarged fragments share a scale and orientation. b, Anions aligned on the metal, V=O bond and long axis. Circles mark the corresponding thione and carbonyl atoms. The first ordered models are shown for dmid and C$_3$S$_5$. Atom colors: C, dark grey; H, light grey; O, red; S, gold; P, orange; V, teal; F, turquoise; Cl, green; Br, brown; I, purple.\par}
\end{figure}
\begin{figure}[!htbp]
\centering\includegraphics[width=\linewidth]{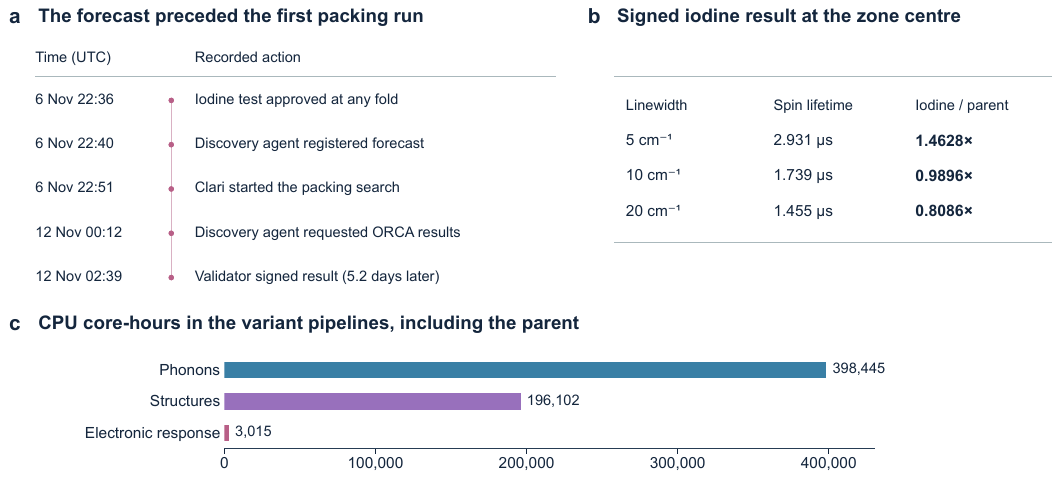}\par
{\raggedright\rmfamily\small \textbf{Supplementary Fig. S3. The iodine calculation and its signed verdict.} a, The forecast preceded the first packing job by 11 minutes and the signed result by 5.2 days (UTC). b, Signed zone-centre lifetimes and iodine-to-parent ratios at 300 K and 0.345 T. c, Allocated CPU core-hours by salt and calculation stage. Table S9 gives campaign totals.\par}
\end{figure}
\begin{figure}[!htbp]
\centering\includegraphics[width=\linewidth]{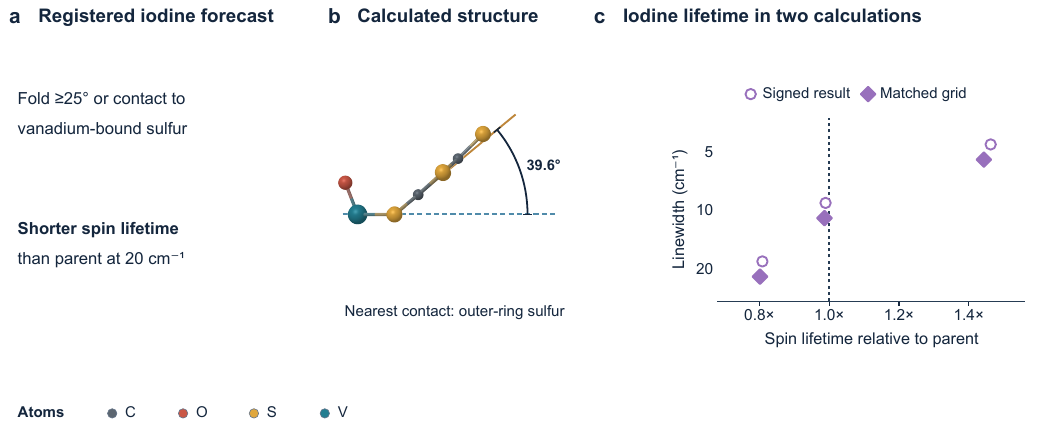}\par
{\raggedright\rmfamily\small \textbf{Supplementary Fig. S4. The registered iodine forecast and its later sampling check.} a, The forecast specified a fold of at least 25\(^{\circ}\) or an iodine contact towards a vanadium-bound sulfur, and a lifetime loss at 20 cm\(^{-1}\). b, Iodine folded by 39.6\(^{\circ}\) and its nearest sulfur was on the outer ring. c, Zone-centre (circles) and matched-grid (diamonds) lifetime ratios at 300 K and 0.345 T. Atom colors: C, dark grey; O, red; S, gold; V, teal.\par}
\end{figure}
\begin{figure}[!htbp]
\centering\includegraphics[width=\linewidth]{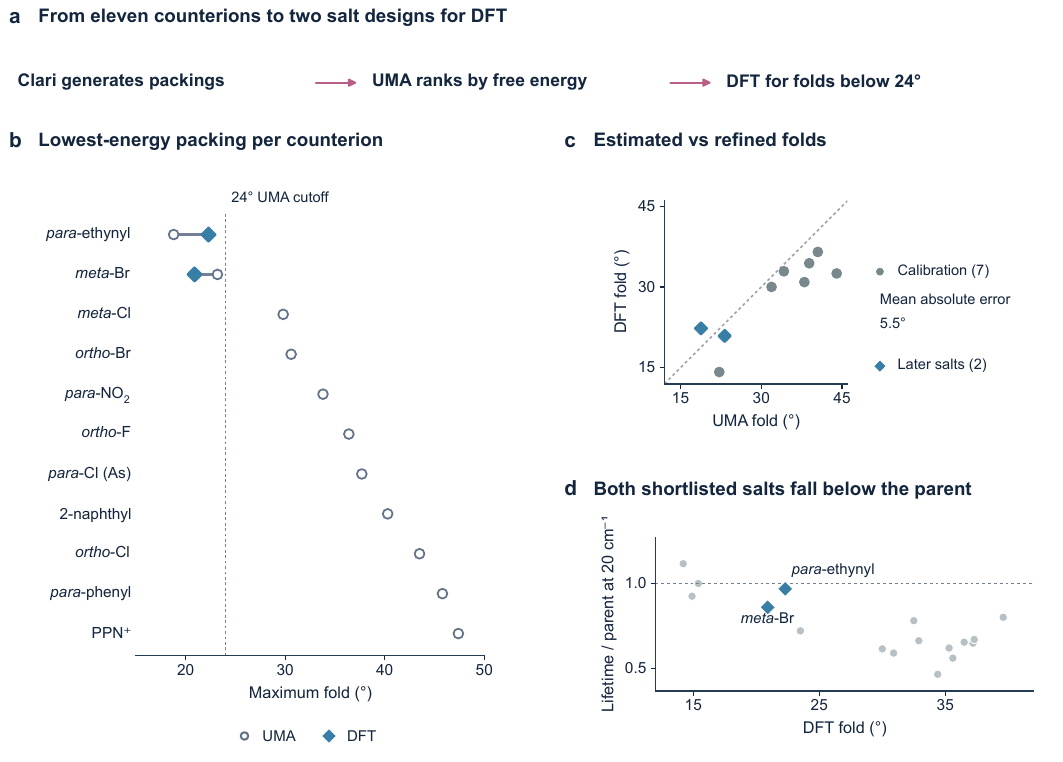}\par
{\raggedright\rmfamily\small \textbf{Supplementary Fig. S5. A 24\(^{\circ}\) packing prescreen selected two counterions for DFT.} a, Clari generated packings for eleven counterions. UMA ranked them by estimated free energy at 300 K. b, Maximum ligand folds from UMA (circles) and DFT (diamonds). The dashed vertical line marks the 24\(^{\circ}\) screening cutoff. c, UMA and DFT folds for seven calibration salts and two later selections (diamonds). The diagonal marks equality. d, Selected salts (diamonds) among all 17 models at \(\sigma\) = 20 cm\(^{-1}\), 300 K and 0.345 T. The dashed line marks the parent.\par}
\end{figure}
\begin{figure}[!htbp]
\centering\includegraphics[width=\linewidth]{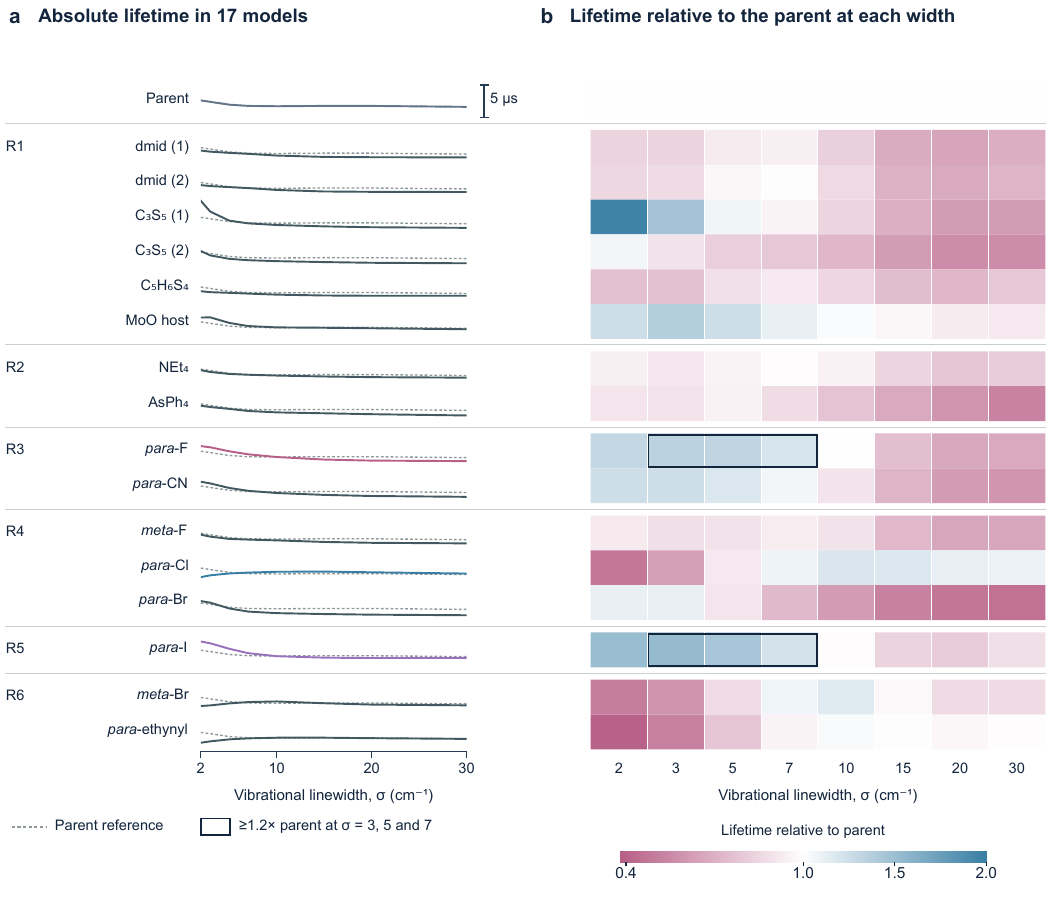}\par
{\raggedright\rmfamily\small \textbf{Supplementary Fig. S6. Absolute and relative lifetimes for each crystal model.} a, Absolute lifetimes with the parent curve dashed and a 5 \(\mu\)s scale bar. b, Lifetime ratios to the parent. Outlines mark models reaching 1.2 times the parent at 3, 5 and 7 cm\(^{-1}\). R1–R6 denote rounds and parentheses distinguish ordered structures. Matched-grid powder averages at 300 K and 0.345 T are tabulated in Tables S3–S4.\par}
\end{figure}
\begin{figure}[!htbp]
\centering\includegraphics[width=\linewidth]{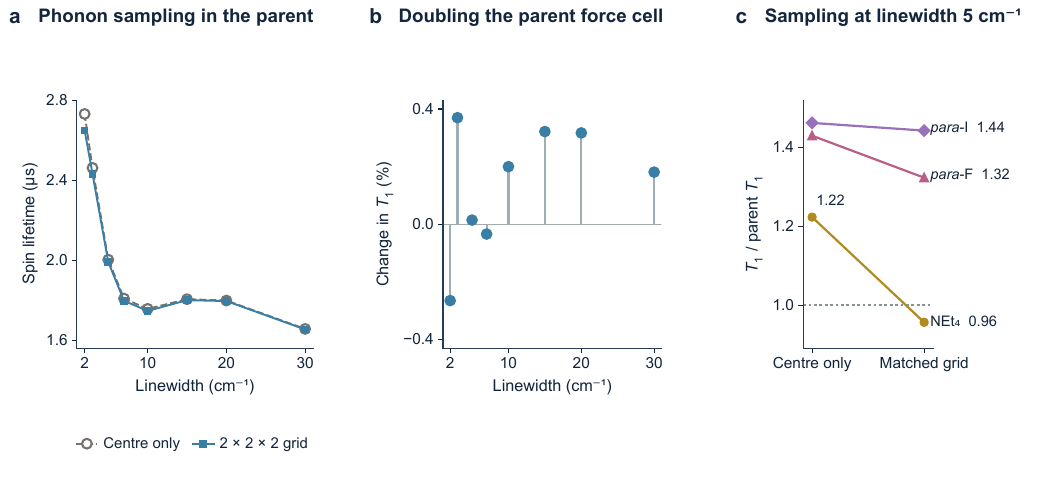}\par
{\raggedright\rmfamily\small \textbf{Supplementary Fig. S7. Sensitivity to phonon sampling and the force-constant supercell.} a, Parent lifetimes at the zone centre and on a 2 \(\times\) 2 \(\times\) 2 grid using primitive-cell forces. b, On the same grid, independent 2 \(\times\) 1 \(\times\) 1 supercell forces changed the parent lifetime by at most 0.37\%. c, At \(\sigma\) = 5 cm\(^{-1}\), denser sampling removed the tetraethylammonium gain. Iodine and fluorine retained theirs. Powder averages at 300 K and 0.345 T.\par}
\end{figure}
\begin{figure}[!htbp]
\centering\includegraphics[width=\linewidth]{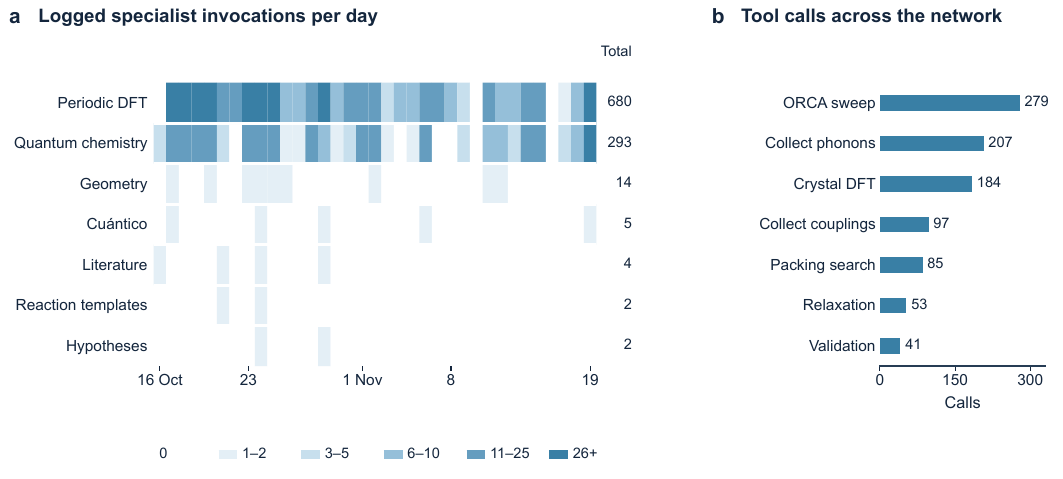}\par
{\raggedright\rmfamily\small \textbf{Supplementary Fig. S8. Specialist activity during the search.} a, Daily specialist invocations attributed to the discovery agent, 16 October to 19 November 2026. Totals are at right. White denotes zero. b, Calls to scientific tools across all agents. An invocation could include several calls.\par}
\end{figure}
\pdfbookmark[1]{Phonon sampling}{secPhononsampling}
\section*{Phonon sampling}
\medskip{\sffamily\small \textbf{Supplementary Table S2. Parent lifetime (\(\mu\)s) by force-constant model and grid.}\par}\nopagebreak
{\centering\small\sffamily\setlength{\tabcolsep}{3.2pt}\renewcommand{\arraystretch}{1.16}\begin{tabular*}{\linewidth}{@{}l@{\extracolsep{\fill}}cccccc@{}}\toprule
\textbf{Force constants} & \textbf{q grid} & \textbf{2} & \textbf{5} & \textbf{10} & \textbf{20} & \textbf{30} \\ \midrule
DFT primitive-cell forces & 1\(\times\)1\(\times\)1 & 2.732 & 2.004 & 1.758 & 1.799 & 1.657 \\
DFT primitive-cell forces & 2\(\times\)2\(\times\)2 & 2.650 & 1.992 & 1.748 & 1.797 & 1.655 \\
DFT supercell forces & 2\(\times\)2\(\times\)2 & 2.643 & 1.992 & 1.751 & 1.802 & 1.658 \\
DFT + UMA dispersion & 1\(\times\)1\(\times\)1 & 2.732 & 2.004 & 1.758 & 1.799 & 1.657 \\
DFT + UMA dispersion & 2\(\times\)2\(\times\)2 & 2.532 & 1.975 & 1.734 & 1.774 & 1.631 \\
UMA supercell forces & 2\(\times\)2\(\times\)2 & 0.470 & 0.255 & 0.314 & 0.391 & 0.422 \\
UMA supercell forces & 3\(\times\)3\(\times\)2 & 0.491 & 0.254 & 0.313 & 0.391 & 0.422 \\
\bottomrule\end{tabular*}\par}
Columns give the pair broadening \(\sigma\) in cm\(^{-1}\). All values refer to 300 K and 0.345 T. On the dense grid, replacing the primitive-cell force constants with independent 2\(\times\)1\(\times\)1 DFT supercell values changed the parent lifetime by at most 0.4\% across all eight sampled widths. With DFT frequencies at \(\Gamma\) and UMA dispersion, moving from \(\Gamma\) to the 2\(\times\)2\(\times\)2 grid changed the lifetime by 1.4--1.6\% at the tabulated points from 5 to 30 cm\(^{-1}\) and by 7.3\% at 2 cm\(^{-1}\). Using UMA force constants alone gave substantially shorter lifetimes. The candidate comparisons all use DFT force constants.
\par
\pdfbookmark[1]{Lifetimes across crystal models}{secLifetimesacrosscrystalmodels}
\section*{Lifetimes across crystal models}
Tables S3 and S4 are the final comparison across all completed models. Table S10 records the results available when the next design was chosen. Each crystal model uses its own DFT force constants and a reciprocal grid with equivalent-supercell widths of at least 21 \AA{}. All lifetimes are powder averages at 0.345 T and 300 K. Numbers in parentheses distinguish alternative ordered structures of the same composition. For substituted phosphonium cations, \textit{para} and \textit{meta} give the position of the substituent on a phenyl ring.
\par
\medskip{\sffamily\small \textbf{Supplementary Table S3. Spin--lattice relaxation time (\(\mu\)s).}\par}\nopagebreak
{\centering\small\sffamily\setlength{\tabcolsep}{3.2pt}\renewcommand{\arraystretch}{1.16}\begin{tabular*}{\linewidth}{@{}l@{\extracolsep{\fill}}cccccccc@{}}\toprule
\textbf{Crystal model} & \textbf{2} & \textbf{3} & \textbf{5} & \textbf{7} & \textbf{10} & \textbf{15} & \textbf{20} & \textbf{30} \\ \midrule
Parent & 2.65 & 2.43 & 1.99 & 1.80 & 1.75 & 1.80 & 1.80 & 1.65 \\
dmid (1) & 2.20 & 2.03 & 1.85 & 1.69 & 1.43 & 1.22 & 1.17 & 1.14 \\
dmid (2) & 2.25 & 2.10 & 1.93 & 1.79 & 1.50 & 1.26 & 1.21 & 1.18 \\
C\textsubscript{3}S\textsubscript{5} (1) & 5.21 & 3.54 & 2.14 & 1.72 & 1.46 & 1.25 & 1.12 & 1.02 \\
C\textsubscript{3}S\textsubscript{5} (2) & 2.83 & 2.15 & 1.62 & 1.41 & 1.26 & 1.12 & 1.01 & 0.93 \\
C\textsubscript{5}H\textsubscript{6}S\textsubscript{4} & 2.01 & 1.85 & 1.75 & 1.63 & 1.47 & 1.34 & 1.30 & 1.30 \\
MoO host & 3.32 & 3.35 & 2.49 & 2.00 & 1.80 & 1.74 & 1.66 & 1.51 \\
NEt\textsubscript{4} & 2.49 & 2.21 & 1.91 & 1.78 & 1.65 & 1.51 & 1.40 & 1.33 \\
AsPh\textsubscript{4} & 2.38 & 2.17 & 1.89 & 1.56 & 1.34 & 1.20 & 1.06 & 0.87 \\
\textit{para}-F & 3.44 & 3.25 & 2.64 & 2.17 & 1.75 & 1.36 & 1.19 & 1.11 \\
\textit{para}-CN & 3.32 & 3.06 & 2.35 & 1.91 & 1.57 & 1.28 & 1.11 & 0.98 \\
\textit{meta}-F & 2.44 & 2.13 & 1.78 & 1.67 & 1.55 & 1.31 & 1.18 & 1.09 \\
\textit{para}-Cl & 1.25 & 1.54 & 1.82 & 1.96 & 2.09 & 2.12 & 2.01 & 1.81 \\
\textit{para}-Br & 2.95 & 2.70 & 1.80 & 1.32 & 1.08 & 0.93 & 0.84 & 0.75 \\
\textit{para}-I & 4.00 & 3.72 & 2.88 & 2.19 & 1.72 & 1.50 & 1.44 & 1.45 \\
\textit{meta}-Br & 1.32 & 1.42 & 1.73 & 1.94 & 2.03 & 1.75 & 1.55 & 1.43 \\
\textit{para}-ethynyl & 1.03 & 1.25 & 1.56 & 1.71 & 1.81 & 1.82 & 1.74 & 1.63 \\
\bottomrule\end{tabular*}\par}
\medskip{\sffamily\small \textbf{Supplementary Table S4. Lifetime relative to the parent at the same broadening.}\par}\nopagebreak
{\centering\small\sffamily\setlength{\tabcolsep}{3.2pt}\renewcommand{\arraystretch}{1.16}\begin{tabular*}{\linewidth}{@{}l@{\extracolsep{\fill}}cccccccc@{}}\toprule
\textbf{Crystal model} & \textbf{2} & \textbf{3} & \textbf{5} & \textbf{7} & \textbf{10} & \textbf{15} & \textbf{20} & \textbf{30} \\ \midrule
Parent & 1.00 & 1.00 & 1.00 & 1.00 & 1.00 & 1.00 & 1.00 & 1.00 \\
dmid (1) & 0.83 & 0.83 & 0.93 & 0.94 & 0.82 & 0.68 & 0.65 & 0.69 \\
dmid (2) & 0.85 & 0.86 & 0.97 & 0.99 & 0.86 & 0.70 & 0.67 & 0.71 \\
C\textsubscript{3}S\textsubscript{5} (1) & 1.96 & 1.46 & 1.08 & 0.95 & 0.84 & 0.69 & 0.62 & 0.62 \\
C\textsubscript{3}S\textsubscript{5} (2) & 1.07 & 0.89 & 0.81 & 0.78 & 0.72 & 0.62 & 0.56 & 0.56 \\
C\textsubscript{5}H\textsubscript{6}S\textsubscript{4} & 0.76 & 0.76 & 0.88 & 0.91 & 0.84 & 0.74 & 0.72 & 0.79 \\
MoO host & 1.25 & 1.38 & 1.25 & 1.11 & 1.03 & 0.96 & 0.93 & 0.91 \\
NEt\textsubscript{4} & 0.94 & 0.91 & 0.96 & 0.99 & 0.95 & 0.84 & 0.78 & 0.80 \\
AsPh\textsubscript{4} & 0.90 & 0.89 & 0.95 & 0.87 & 0.77 & 0.67 & 0.59 & 0.52 \\
\textit{para}-F & 1.30 & 1.34 & 1.32 & 1.21 & 1.00 & 0.75 & 0.66 & 0.67 \\
\textit{para}-CN & 1.25 & 1.26 & 1.18 & 1.06 & 0.90 & 0.71 & 0.62 & 0.59 \\
\textit{meta}-F & 0.92 & 0.88 & 0.89 & 0.93 & 0.89 & 0.73 & 0.66 & 0.66 \\
\textit{para}-Cl & 0.47 & 0.63 & 0.92 & 1.09 & 1.19 & 1.18 & 1.12 & 1.09 \\
\textit{para}-Br & 1.11 & 1.11 & 0.90 & 0.73 & 0.62 & 0.52 & 0.47 & 0.45 \\
\textit{para}-I & 1.51 & 1.53 & 1.44 & 1.22 & 0.99 & 0.83 & 0.80 & 0.88 \\
\textit{meta}-Br & 0.50 & 0.58 & 0.87 & 1.08 & 1.16 & 0.97 & 0.86 & 0.86 \\
\textit{para}-ethynyl & 0.39 & 0.51 & 0.78 & 0.95 & 1.04 & 1.01 & 0.97 & 0.99 \\
\bottomrule\end{tabular*}\par}
Column headings in Tables S3 and S4 give \(\sigma\) in cm\(^{-1}\). We calculated ratios before rounding the lifetimes.
\par
\pdfbookmark[1]{Packing and electronic structure}{secPackingandelectronicstructure}
\section*{Packing and electronic structure}
\medskip{\sffamily\small \textbf{Supplementary Table S5. Geometry, upper d\(\pi\) energy and dense sampling.}\par}\nopagebreak
{\centering\small\sffamily\setlength{\tabcolsep}{3.2pt}\renewcommand{\arraystretch}{1.16}\begin{tabular*}{\linewidth}{@{}l@{\extracolsep{\fill}}cccccc@{}}\toprule
\textbf{Crystal model} & \textbf{Atoms} & \textbf{V sites} & \textbf{Folds (\(^{\circ}\))} & \textbf{Upper d\(\pi\) (cm\(^{-1}\))} & \textbf{q grid} & \textbf{Modes} \\ \midrule
Parent & 216 & 2 & 15.4 / 15.3 & 24,017 & 2\(\times\)2\(\times\)2 & 1,552 \\
dmid (1) & 216 & 2 & 37.2 / 7.3 & 22,577 & 2\(\times\)2\(\times\)2 & 1,531 \\
dmid (2) & 216 & 2 & 37.3 / 7.7 & 22,543 & 2\(\times\)2\(\times\)2 & 1,527 \\
C\textsubscript{3}S\textsubscript{5} (1) & 216 & 2 & 35.3 / 11.5 & 23,760 & 2\(\times\)2\(\times\)2 & 1,596 \\
C\textsubscript{3}S\textsubscript{5} (2) & 216 & 2 & 35.6 / 12.0 & 23,747 & 2\(\times\)2\(\times\)2 & 1,600 \\
C\textsubscript{5}H\textsubscript{6}S\textsubscript{4} & 268 & 2 & 23.5 / 22.7 & 24,685 & 3\(\times\)2\(\times\)1 & 1,400 \\
MoO host & 216 & 1 & 14.9 / 14.9 & 23,963 & 2\(\times\)2\(\times\)2 & 1,561 \\
NEt\textsubscript{4} & 76 & 1 & 13.8 / 32.5 & 22,617 & 2\(\times\)3\(\times\)3 & 1,399 \\
AsPh\textsubscript{4} & 216 & 2 & 14.0 / 30.9 & 23,320 & 2\(\times\)2\(\times\)2 & 1,653 \\
\textit{para}-F & 108 & 1 & 32.9 / 14.2 & 22,539 & 3\(\times\)3\(\times\)2 & 1,816 \\
\textit{para}-CN & 220 & 2 & 30.0 / 9.2 & 23,019 & 2\(\times\)3\(\times\)2 & 2,479 \\
\textit{meta}-F & 216 & 2 & 11.8 / 36.5 & 22,364 & 2\(\times\)3\(\times\)2 & 2,450 \\
\textit{para}-Cl & 432 & 4 & 2.8 / 14.2 & 24,542 & 2\(\times\)2\(\times\)2 & 3,343 \\
\textit{para}-Br & 216 & 2 & 34.4 / 9.2 & 23,111 & 2\(\times\)2\(\times\)2 & 1,652 \\
\textit{para}-I & 108 & 1 & 39.6 / 6.5 & 22,142 & 3\(\times\)2\(\times\)2 & 1,237 \\
\textit{meta}-Br & 432 & 4 & 19.5 / 20.9 & 23,955 & 2\(\times\)1\(\times\)2 & 1,677 \\
\textit{para}-ethynyl & 448 & 4 & 16.4 / 22.3 & 24,476 & 1\(\times\)2\(\times\)2 & 1,649 \\
\bottomrule\end{tabular*}\par}
The V-site column counts spin-bearing vanadyl anions in the calculation cell. The two fold angles refer to the ligands of the anion used for the electronic calculation. The MoO host contains one vanadyl and one molybdenyl site. The abbreviations dmit and dmid denote the thione and carbonyl forms of the ligand. The C\textsubscript{3}S\textsubscript{5} ligand is a connectivity isomer of dmit, and C\textsubscript{5}H\textsubscript{6}S\textsubscript{4} differs in ring composition. Crystallography Open Database entries are 4130644 (parent) \href{https://doi.org/10.1021/jacs.6b05574}{[3]}, 4130985 (dmid) \href{https://doi.org/10.1021/jacs.6b08467}{[5]}, 4130986 (C\textsubscript{3}S\textsubscript{5} isomer) \href{https://doi.org/10.1021/jacs.6b08467}{[5]}, 4125474 (C\textsubscript{5}H\textsubscript{6}S\textsubscript{4}) \href{https://doi.org/10.1021/jacs.6b13030}{[87]} and 4130645 (host) \href{https://doi.org/10.1021/jacs.6b05574}{[3]}.
\par
In the fixed parent environment the upper d\(\pi\) energy was 24,017 cm\(^{-1}\) at the parent's 15.4\(^{\circ}\) maximum fold, 23,504 cm\(^{-1}\) at 30\(^{\circ}\) and 23,195 cm\(^{-1}\) at 35\(^{\circ}\). Replacing only the parent geometry with the \textit{para}-Cl geometry raised this level by 253 cm\(^{-1}\). Replacing only the electrostatic embedding raised it by 300 cm\(^{-1}\). Changing both gave a 525 cm\(^{-1}\) increase.
\par
\pdfbookmark[1]{Rate decomposition and screening}{secRatedecompositionandscreening}
\section*{Rate decomposition and screening}
\medskip{\sffamily\small \textbf{Supplementary Table S6. Spearman correlations with the paired lifetime ratio.}\par}\nopagebreak
{\centering\small\sffamily\setlength{\tabcolsep}{3.2pt}\renewcommand{\arraystretch}{1.16}\begin{tabular*}{\linewidth}{@{}l@{\extracolsep{\fill}}rr@{}}\toprule
\textbf{Descriptor} & \textbf{\(\sigma\) = 3 cm\(^{-1}\)} & \textbf{\(\sigma\) = 20 cm\(^{-1}\)} \\ \midrule
V-bearing anions per cell & \ensuremath{-}0.77 & +0.15 \\
Maximum ligand fold & +0.26 & \ensuremath{-}0.57 \\
Upper d\(\pi\) energy & \ensuremath{-}0.41 & +0.42 \\
Zone-centre share within 10 cm\(^{-1}\) at \(\sigma\) = 20 cm\(^{-1}\) & \ensuremath{-}0.34 & +0.79 \\
\bottomrule\end{tabular*}\par}
Each correlation uses 17 crystal models and lifetimes from matched grids. In the supplied \textit{para}-Cl, \textit{meta}-Br and \textit{para}-ethynyl cells, translating the atoms by $(1/2,1/2,0)$, $(1/2,1/2,1/2)$ and $(1/2,1/2,0)$ in fractional coordinates, respectively, gives a one-to-one match within each chemical element. The largest heavy-atom separations from the translated partners are 0.065, 0.036 and 0.010 Å. As a sensitivity test, assigning two sites to each of these cells gives $\rho=-0.57$ at 3 cm\(^{-1}\) with the original lifetimes. This calculation tests the dependence on cell representation. Exact translational symmetry would require coincident sites. For the rate fraction in the last row, zone-centre decompositions cover all 17 models. Dense-grid decompositions are available for the parent, \textit{para}-Cl, \textit{para}-F and \textit{para}-I (Fig. 6c).
\par
\medskip{\sffamily\small \textbf{Supplementary Table S7. Interference through the upper-d\(\pi\) doublet.}\par}\nopagebreak
{\centering\small\sffamily\setlength{\tabcolsep}{3.2pt}\renewcommand{\arraystretch}{1.16}\begin{tabular*}{\linewidth}{@{}l@{\extracolsep{\fill}}ccc@{}}\toprule
\textbf{Model} & \textbf{Excitation (cm\(^{-1}\))} & \textbf{Only this doublet / full} & \textbf{Without this doublet / full} \\ \midrule
Parent & 24,036 & 34.4 & 41.9 \\
\textit{para}-Cl & 24,550 & 3.4 & 6.0 \\
\textit{para}-Br & 23,135 & 86.4 & 98.2 \\
\textit{para}-I & 22,184 & 183.8 & 202.7 \\
\textit{para}-F & 22,611 & 351.7 & 374.4 \\
\bottomrule\end{tabular*}\par}
The last two columns divide the rate calculated with or without the indicated doublet by the full rate for the same salt at \(\sigma\) = 10 cm\(^{-1}\). Amplitudes through different doublets interfere, so the rate ratios do not add to one. Table S7 lists excited Kramers doublets after spin--orbit coupling. Table S5 lists the orbital levels.
\par
The packing screen selected counterions whose lowest-energy UMA structure had a maximum fold below 24\(^{\circ}\). For seven earlier salts the mean absolute difference between UMA and DFT folds was 5.5\(^{\circ}\), with a maximum of 11.5\(^{\circ}\). The two round-6 candidates were outside this calibration set. In round 5, the chlorinated salts could proceed to phonons only with both folds at or below 20\(^{\circ}\) and a third d--d excitation of at least 23,700 cm\(^{-1}\). The registration numbered the NEVPT2 excitations by increasing energy and required an orbital-character check if the levels reordered. Iodine received the full calculation whatever its fold. In round 6, before the outcomes were known, the investigator designated the upper d\(\pi\) level as the criterion for proceeding to phonons.
\par
\pdfbookmark[1]{Computational resources}{secComputationalresources}
\section*{Computational resources}
\medskip{\sffamily\small \textbf{Supplementary Table S8. Allocated CPU core-hours by stage and candidate.}\par}\nopagebreak
{\centering\small\sffamily\setlength{\tabcolsep}{3.2pt}\renewcommand{\arraystretch}{1.16}\begin{tabular*}{\linewidth}{@{}l@{\extracolsep{\fill}}ccccc@{}}\toprule
\textbf{Candidate} & \textbf{DFT relaxation} & \textbf{DFT phonons} & \textbf{ORCA} & \textbf{Other} & \textbf{Total\textsuperscript{1}} \\ \midrule
Parent (in-loop) & 0 & 23,364 & 221 & 1 & 23,862 \\
dmid & 2,588 & 9,253 & 178 & 0 & 12,019 \\
C\textsubscript{3}S\textsubscript{5} & 49,843 & 9,946 & 209 & 0 & 59,998 \\
C\textsubscript{5}H\textsubscript{6}S\textsubscript{4} & 8,721 & 9,845 & 274 & 0 & 18,841 \\
MoO host & 2,146 & 6,265 & 100 & 0 & 8,510 \\
NEt\textsubscript{4} & 19,939 & 5,124 & 119 & 0 & 25,182 \\
AsPh\textsubscript{4} & 26,885 & 8,843 & 173 & 0 & 35,900 \\
\textit{para}-F & 2,706 & 2,090 & 185 & 0 & 4,982 \\
\textit{para}-CN & 16,700 & 9,879 & 191 & 0 & 26,770 \\
\textit{meta}-F & 15,994 & 13,985 & 317 & 0 & 30,296 \\
\textit{para}-Cl & 10,914 & 89,235 & 277 & 0 & 100,427 \\
\textit{para}-Br & 11,444 & 10,908 & 153 & 0 & 22,505 \\
Two chloroaryls & 6,358 & 0 & 0 & 0 & 6,358 \\
Four chloroaryls & 3,545 & 0 & 1 & 0 & 3,546 \\
\textit{para}-I & 6,101 & 5,371 & 155 & 0 & 11,628 \\
\textit{meta}-Br & 3,478 & 105,620 & 274 & 0 & 109,372 \\
\textit{para}-ethynyl & 8,740 & 88,717 & 188 & 0 & 97,644 \\
Coordinator studies & 115 & 63,013 & 0 & 510 & 63,642 \\
\bottomrule\end{tabular*}\par}
\textsuperscript{1} The scheduler recorded a total for each candidate, combining its alternative ordered structures. Some totals differ slightly from the sums of the displayed stage entries. The coordinator row includes work outside candidate production. Job accounting grouped the GPU packing predictions for several salts under the parent.
\par
\medskip{\sffamily\small \textbf{Supplementary Table S9. Campaign accounting at each reporting scope.}\par}\nopagebreak
{\centering\small\sffamily\setlength{\tabcolsep}{3.2pt}\renewcommand{\arraystretch}{1.16}\begin{tabular*}{\linewidth}{@{}l@{\extracolsep{\fill}}rr@{}}\toprule
\textbf{Scope} & \textbf{CPU core-h} & \textbf{GPU-h} \\ \midrule
Variant pipelines & 597,840 & 69.1 \\
Coordinator studies & 63,642 & 1.0 \\
Pre-launch parent baseline & 11,937 & --- \\
Pre-DFT counterion screen & --- & 70.9 \\
Total campaign use & 673,419 & 141.0 \\
\bottomrule\end{tabular*}\par}
The variant total sums the scheduler totals in Table S8. Of these core-hours, 597,563 were assigned to individual calculation stages. The run-history appendix records the estimate available during the campaign. The structure search generated 6,798 proposed packings across 22 counterions, followed by 114 DFT relaxation runs. Phonons dominated the CPU time for the largest cells. The 432- and 448-atom round-6 cells required 1,296 and 1,344 single-coordinate displacements, respectively.
\par
\pdfbookmark[1]{Agent activity and decisions}{secAgentactivityanddecisions}
\section*{Agent activity and decisions}
Cuántico in turn consulted La Agente Teórica on 6 and 19 November to check the mathematics. The job driver started or resumed work when the queue changed. The agent logs contain 1,365 invocations and 18,028 tool calls, totalling 1.87 billion input tokens (including cached prefixes) and 12.0 million output tokens. The campaign ledger records 280 orchestrator turns, while the agent log records 281 invocations for that role. For periodic DFT, the driver recorded 403 delegations and the specialist thread log recorded 680 invocations (Supplementary Fig. S8a). The scheduler separately recorded 13,712 computational jobs.
\par
For each submitted lifetime calculation, the validator received the proposed edit, the parent snapshot and paths to the couplings, displacement manifest and phonons. It recomputed the lifetime, compared the vibrational mode families and issued a receipt. The receipts for \textit{para}-I and \textit{para}-F are rcpt-f0670a4fa30c and rcpt-d3d6401285af, respectively. Table S10 traces why fluorine led to a new set of designs and why iodine was tested. Table S4 gives the later comparison that put both salts on the shortlist.
\par
Seven annotated exchanges in the accompanying data show how a proposal, job request or computed result passed between agents. The appendix reproduces the campaign run history, with its original figures and transcripts formatted for print. The supplied HTML and PDF accompany the source data.
\par
\pdfbookmark[1]{Decisions leading to the iodine test}{secDecisionsleadingtotheiodinetest}
\section*{Decisions leading to the iodine test}
\medskip{\sffamily\small \textbf{Supplementary Table S10. How each round informed the next chemical test.} Parentheses give the number of salt designs tested. Two round-1 salts each had two ordered structures.\par}\nopagebreak
{\centering\small\sffamily\renewcommand{\arraystretch}{1.08}\begin{tabular}{@{}p{10mm}@{\hspace{3mm}}p{39mm}@{\hspace{3mm}}p{56mm}@{\hspace{3mm}}p{54mm}@{}}\toprule
\textbf{Round} & \textbf{Hypothesis} & \textbf{Calculated outcome} & \textbf{Next decision or analysis} \\ \midrule
{\raggedright 1 (4)\par} & {\raggedright Changing the dithiolene should weaken the V--S stretch and V=O tilt.\par} & {\raggedright The carbonyl edit suppressed the target motion, but other vibrations gained more relaxation rate. The host control kept the original ligand nearly flat.\par} & {\raggedright Retain the spin-bearing anion and change its environment.\par} \\[3pt]
{\raggedright 2 (2)\par} & {\raggedright A counterion swap should constrain donor motion or change vibrational mixing.\par} & {\raggedright Tetraethylammonium and tetraphenylarsonium both gave folds above 30\(^{\circ}\) and shorter broad-linewidth lifetimes.\par} & {\raggedright Keep the phosphonium core and replace one aryl hydrogen.\par} \\[3pt]
{\raggedright 3 (2)\par} & {\raggedright \textit{Para}-F should reduce relaxation through other vibrations. \textit{Para}-CN should pin the anion.\par} & {\raggedright Other vibrations gained 39,000 s\(^{-1}\) in \textit{para}-F at \(\sigma\) = 10 cm\(^{-1}\), opposite to the forecast. At the zone centre, the lifetime reached 1.06 times the parent value for \textit{para}-F and 0.88 for \textit{para}-CN.\par} & {\raggedright The investigator extended the search. Test position with \textit{meta}-F and halogen identity with \textit{para}-Cl and \textit{para}-Br.\par} \\[3pt]
{\raggedright 4 (3)\par} & {\raggedright \textit{Para}-Cl should give modest relief. \textit{Para}-Br should be a failure control.\par} & {\raggedright Cl faced an outer-ring sulfur (14.2\(^{\circ}\) maximum fold). Br contacted a V-bound sulfur (34.4\(^{\circ}\)). The more folded crystal relaxed faster at broad linewidth.\par} & {\raggedright Test whether folding lowers the upper d\(\pi\) energy, then predict a broad-linewidth loss for \textit{para}-I.\par} \\[3pt]
{\raggedright 5 (3)\par} & {\raggedright \textit{Para}-I should fold or contact a V-bound sulfur and lose at \(\sigma\) = 20 cm\(^{-1}\). Additional chloroaryl groups should preserve flat ligands.\par} & {\raggedright \textit{Para}-I folded by 39.6\(^{\circ}\). Its zone-centre lifetime ratio was 0.81 at 20 cm\(^{-1}\) as predicted, and unexpectedly 1.46 at 5 cm\(^{-1}\). Both chlorine salts failed the fold test. The third excitation in \textit{para}-Cl\(_4\) was 23,227 cm\(^{-1}\), below 23,700 cm\(^{-1}\).\par} & {\raggedright In parallel, screen new counterions for flat packings. Two advanced on 8 November, four days before the iodine verdict.\par} \\[3pt]
{\raggedright 6 (2)\par} & {\raggedright Maximum fold \(\leq\)20\(^{\circ}\), upper d\(\pi\) energy \(\geq\)23,700 cm\(^{-1}\) and lifetime at least equal to the parent at \(\sigma\) = 10 and 20 cm\(^{-1}\).\par} & {\raggedright Both cleared the electronic gate. \textit{Meta}-Br and \textit{para}-ethynyl folded by 20.9\(^{\circ}\) and 22.3\(^{\circ}\). The zone-centre results exceeded the parent at 10 cm\(^{-1}\) and fell short at 20 cm\(^{-1}\).\par} & {\raggedright Compare all completed models across linewidths.\par} \\[3pt]
\bottomrule\end{tabular}\par}\vspace{3pt}
On 29 October the search agent sent the hypothesis specialist the approved \textit{meta}-F, \textit{para}-Cl and \textit{para}-Br tests, identifying \textit{para}-Br as the predicted failure. The specialist supplied mechanisms, numerical forecasts and criteria for falsification before the calculations began. The discovery agent chose iodine to test the trend seen with bromine. The investigator authorized the three round-5 candidates for execution and specified that only the chlorinated salts could stop at the structural screen. The discovery agent recorded the forecasts against the original parent.
\par
The \href{../preregistration/round5_hypotheses_ORIGINAL.md}{original round-5 forecast} was written at 22:40:55 UTC on 6 November, before the first packing calculation at 22:51:51 UTC and 5.2 days before the \href{../preregistration/judge_receipt_R5_pI_rcpt-f0670a4fa30c.json}{validator signed the iodine result} at 02:39:18 UTC on 12 November (see the \href{../preregistration/PROVENANCE.md}{timeline in the accompanying preregistration folder}). The corresponding agent exchange was logged at 02:39:34 UTC (run-history appendix). The receipt reports the iodine lifetime. The round ledger records the structural failures of both chlorinated designs. The \textit{para}-Cl\(_4\) gate record also gives its third excitation energy.
\par
The validator could not track the parent vibrational family into \textit{para}-I. Its receipt therefore establishes the total lifetime comparison, while the microscopic assignment remains open. Supplementary Fig. S4c compares the signed result with the later matched-grid calculation.
\par
\par
\pdfbookmark[1]{Supplementary Note 1. From periodic forces to the coupling of a local spin}{suppmath1}
\section*{Supplementary Note 1. From periodic forces to the coupling of a local spin}
\renewcommand{\theequation}{S\arabic{equation}}
Let $u_{l\kappa a}$ be the displacement along Cartesian axis $a$ of atom $\kappa$ in cell $l$, and let $F_{l\kappa a}$ be the corresponding force. The force constant linking this coordinate in cell zero to coordinate $b$ of atom $\tau$ in cell $l$ is $\Phi_{0\kappa a,l\tau b}=-\partial F_{0\kappa a}/\partial u_{l\tau b}$. Each displaced DFT calculation provides forces on every atom. With atomic masses $M_\kappa$ and equilibrium positions $\mathbf r_{l\kappa}$, the Phonopy phase convention \href{https://doi.org/10.1016/j.scriptamat.2015.07.021}{[94]} gives
\par
\begin{equation}
D_{\kappa a,\tau b}(\mathbf q)=
\frac{\sum_l\Phi_{0\kappa a,l\tau b}\,
\exp[i\mathbf q\cdot(\mathbf r_{l\tau}-\mathbf r_{0\kappa})]}
{\sqrt{M_\kappa M_{\tau}}},
\qquad
D(\mathbf q)\mathbf e_{qj}=\omega_{qj}^2\mathbf e_{qj}.
\label{suppmath:eq1}
\end{equation}
Here $j$ labels a branch, and the polarization vector $\mathbf e_{qj}$ has unit norm within the cell used to construct the force constants. With $N$ cells in the sampling supercell, the mass-weighted pattern on the full supercell is
\par
\begin{equation}
p_{l\kappa a}(\mathbf qj)=N^{-1/2}\,
e_{\kappa a}(\mathbf qj)\exp(i\mathbf q\cdot\mathbf r_{l\kappa}).
\label{suppmath:eq2}
\end{equation}
The factor $N^{-1/2}$ keeps the sum of squared amplitudes equal to one when the sampling grid changes. If $2\mathbf q$ is a reciprocal-lattice vector, the physical pattern can be chosen real. For each remaining pair $\{\mathbf q,-\mathbf q\}$, $\sqrt{2}\,\mathrm{Re}\,\mathbf p$ and $\sqrt{2}\,\mathrm{Im}\,\mathbf p$ give two real, normalized standing waves. We count each such pair of wavevectors once. In this real basis a scalar mass-weighted normal coordinate $Q_\lambda$ corresponds to a Cartesian displacement $u_{l\kappa a}=p_{l\kappa a,\lambda}Q_\lambda/\sqrt{M_\kappa}$.
\par
Only the components of that pattern on the embedded complex enter the local Hamiltonian derivative. Let $\mathbf d_k$ be the mass-weighted direction of electronic displacement $k$. Express the projection of the local mode onto the sampled displacement space as $\sum_k c_{\lambda k}\mathbf d_k$. By linearity of the directional derivative, its coupling is obtained from the matrices
\par
\begin{equation}
B_k\simeq\frac{H_{\rm el}(+s_k)-H_{\rm el}(-s_k)}{2s_k},
\qquad
\frac{\partial H_{\rm el}}{\partial Q_\lambda}\simeq\sum_k c_{\lambda k}B_k .
\label{suppmath:eq3}
\end{equation}
Here $s_k$ is the step along $\mathbf d_k$ in mass-weighted coordinates. The positive and negative Hamiltonians must refer to the same orbital frame for their difference to describe a physical derivative. A typical embedded calculation used about 51 displacement directions. Reusing these derivatives decouples the number of electronic calculations from the number of phonons on the final grid. The latter is fixed by the crystal force constants, cell dimensions and vibrational range. The parent used a $2\times2\times2$ grid, \textit{para}-I a $3\times2\times2$ grid and \textit{para}-F a $3\times3\times2$ grid. Table S2 compares changes in the wavevector grid and in the force-constant supercell separately.
\par
\pdfbookmark[1]{Supplementary Note 2. How phonon pairs and electronic pathways enter the rate}{suppmath2}
\section*{Supplementary Note 2. How phonon pairs and electronic pathways enter the rate}
For a given field direction, let $p_f$ and $p_i$ be the populations of the two ground states. With $p_f+p_i=1$, their rate equation gives
\par
\begin{equation}
\dot p_f=W_{f\leftarrow i}p_i-W_{i\leftarrow f}p_f
=-\big(W_{f\leftarrow i}+W_{i\leftarrow f}\big)(p_f-p_f^{\rm eq}).
\label{suppmath:eq4}
\end{equation}
The inverse relaxation time therefore contains both directions of population transfer. At 300 K the thermal energy greatly exceeds the Zeeman splitting at 0.345 T. In that limit the two rates approach one another, giving the familiar factor of two relative to a single spin-flip rate \href{https://doi.org/10.1126/sciadv.adr0168}{[116]}.
\par
For each field direction $\hat{\mathbf n}_k$, we diagonalized the ion Hamiltonian in that field and calculated the longitudinal rate $R_k=W_{f\leftarrow i}+W_{i\leftarrow f}$ by summing the phonon-pair contributions. On the 14-point Lebedev grid, each of the six Cartesian axis directions has weight $w_k=1/15$. Each of the eight directions $(\pm1,\pm1,\pm1)/\sqrt{3}$ has weight $w_k=3/40$. These weights sum to unity. The reported powder lifetime is
\[
T_{1,\mathrm{powder}}=\left(\sum_{k=1}^{14}w_k R_k\right)^{-1}.
\]
This is the weighted harmonic mean of the directional lifetimes. The campaign calculation, the independent validator and the dense-grid linewidth scan all used the same routine to average the rates.
\par
Distinct ordered phonon pairs contribute additively to the transition rate. Write $K_{\alpha\beta}(\sigma)$ for the longitudinal rate contribution with phonon $\alpha$ absorbed and $\beta$ emitted. For a wavenumber threshold $d$, the reported mismatch fraction is
\par
\begin{equation}
F_\sigma(d)=\frac{\sum_{\alpha\ne\beta}K_{\alpha\beta}(\sigma)\,
\mathbf 1\{|\tilde\nu_\alpha-\tilde\nu_\beta|\le d\}}
{\sum_{\alpha\ne\beta}K_{\alpha\beta}(\sigma)},
\qquad \tilde\nu_\lambda=\epsilon_\lambda/(hc).
\label{suppmath:eq5}
\end{equation}
The indicator selects pairs by their frequency difference. The rate from pairs separated by more than $d$ is $[1-F_\sigma(d)]/T_1$, where $T_1$ and the pair fractions use the same linewidth and sampling. The Gaussian in the rate itself also includes the spin splitting. For example, among the leading matched-grid pairs for \textit{para}-I at $\sigma=20$ cm\(^{-1}\) are modes at 227.5 and 209.8 cm\(^{-1}\). Absorbing the higher-frequency mode and emitting the lower one leaves a mismatch of about 17.4 cm\(^{-1}\) for a spin excitation of 0.3 cm\(^{-1}\). At that offset the Gaussian is about 69\% of its peak height for $\sigma=20$ cm\(^{-1}\) and about $5\times10^{-8}$ of its peak for $\sigma=3$ cm\(^{-1}\). These are relative line-shape factors. The actual fraction of the total rate also depends on the electronic matrix elements, occupations and all competing pairs.
\par
Within one phonon pair, paths through different excited doublets lead to the same final state. If $A_m$ denotes the amplitude summed over the two components of excited doublet $m$ and both time orderings, then
\par
\begin{equation}
\left|\sum_m A_m\right|^2=\sum_m|A_m|^2
+2\sum_{m<k}\mathrm{Re}(A_m A_k^*).
\label{suppmath:eq6}
\end{equation}
Restricting the amplitude to one doublet removes its interference with the others. In \textit{para}-I at $\sigma=10$ cm\(^{-1}\), retaining only the upper-$d\pi$ doublet gives 183.8 times the full rate. Removing that doublet gives 202.7 times the full rate (Table S7). Both exceed one because the omitted cross terms are destructive. A pathway ratio consequently has a different interpretation from the phonon-pair fraction $F_\sigma(d)$, which lies between zero and one.
\par
\clearpage\newgeometry{left=3mm,right=3mm,top=3mm,bottom=3mm}\pagestyle{empty}
\pdfbookmark[1]{Campaign run history}{campaignrunhistory}
\noindent\includegraphics[page=1,width=\textwidth,height=.992\textheight,keepaspectratio]{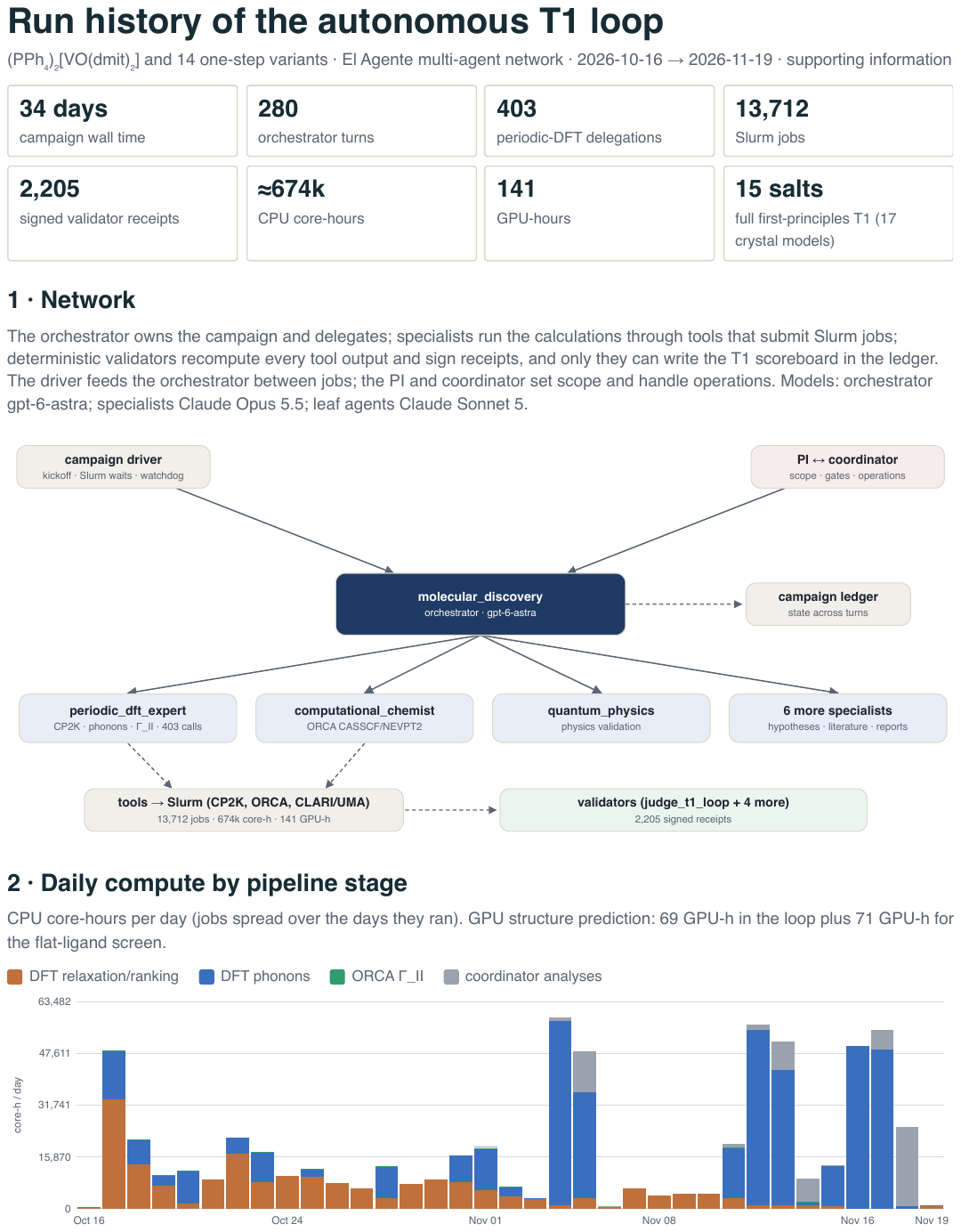}\par
\newpage
\noindent\includegraphics[page=2,width=\textwidth,height=.992\textheight,keepaspectratio]{appendix/run_history_print.pdf}\par
\newpage
\noindent\includegraphics[page=3,width=\textwidth,height=.992\textheight,keepaspectratio]{appendix/run_history_print.pdf}\par
\newpage
\noindent\includegraphics[page=4,width=\textwidth,height=.992\textheight,keepaspectratio]{appendix/run_history_print.pdf}\par
\newpage
\noindent\includegraphics[page=5,width=\textwidth,height=.992\textheight,keepaspectratio]{appendix/run_history_print.pdf}\par
\newpage
\noindent\includegraphics[page=6,width=\textwidth,height=.992\textheight,keepaspectratio]{appendix/run_history_print.pdf}\par
\newpage
\noindent\includegraphics[page=7,width=\textwidth,height=.992\textheight,keepaspectratio]{appendix/run_history_print.pdf}\par
\newpage
\noindent\includegraphics[page=8,width=\textwidth,height=.992\textheight,keepaspectratio]{appendix/run_history_print.pdf}\par
\newpage
\noindent\includegraphics[page=9,width=\textwidth,height=.992\textheight,keepaspectratio]{appendix/run_history_print.pdf}\par
\newpage
\noindent\includegraphics[page=10,width=\textwidth,height=.992\textheight,keepaspectratio]{appendix/run_history_print.pdf}\par
\newpage
\noindent\includegraphics[page=11,width=\textwidth,height=.992\textheight,keepaspectratio]{appendix/run_history_print.pdf}\par

\end{document}